\documentclass[aps,reprint,superscriptaddress,nofootinbib]{revtex4-1} 
\usepackage[english]{babel}

\usepackage{amsmath, amssymb}
\usepackage{mathtools}
\usepackage{siunitx}
\usepackage{environ}
\usepackage{mathrsfs}
\usepackage{braket}
\usepackage{bm}
\DeclareMathAlphabet{\mathscrlower}{OT1}{pzc}{m}{it} 
\usepackage{prettyref}
\usepackage{graphicx}
\usepackage{booktabs}
\usepackage{array}
\usepackage{multirow}
\usepackage{dcolumn}
\usepackage{threeparttable}
\usepackage{threeparttablex}
\usepackage{placeins}
\usepackage[stable]{footmisc}

\newcommand{\pauli}{\boldsymbol{\sigma}}
\newcommand{\Pauli}{\boldsymbol{\sigma}}
\newcommand{\diraccontra}[1]{\boldsymbol{\gamma}^{#1}}

\newcommand{\diraca}{\vec{\boldsymbol{\alpha}}}

\newcommand{\unity}{\bm{1}_{2\times 2}}
\newcommand{\zeroty}{\bm{0}_{2\times 2}}
\newcommand{\pos}{\vec{r}}

\newcommand{\momop}{\hat{\vec{p}}}
\newcommand{\momoppot}{\hat{\vec{\pi}}}
\newcommand{\spinmom}{\vec{\Pauli}\cdot\momop}
\newcommand{\spinmompot}{\vec{\Pauli}\cdot\momoppot}
\newcommand{\efield}{\mathcal{E}}
\newcommand{\bfield}{\mathcal{B}}
\newcommand{\Sum}[2]{\sum\limits_{#1}^{#2}}
\newcommand{\parantheses}[1]{\left(#1\right)}
\newcommand{\brackets}[1]{\left[#1\right]}
\newcommand{\braces}[1]{\left\{ #1\right\}}
\let\nablatmp\nabla
\renewcommand{\nabla}{\vec{\nablatmp}}
\DeclarePairedDelimiter\abs{\lvert}{\rvert}
\makeatletter
\let\oldabs\abs
\def\abs{\@ifstar{\oldabs}{\oldabs*}}

\newcommand{\Op}[1]{\hat{#1}}

\AtBeginDocument{
\heavyrulewidth=.08em
\lightrulewidth=.05em
\cmidrulewidth=.03em
\belowrulesep=.65ex
\belowbottomsep=0pt
\aboverulesep=.4ex
\abovetopsep=0pt
\cmidrulesep=\doublerulesep
\cmidrulekern=.5em
\defaultaddspace=.5em
}
\begin{document}
\title{Systematic study of relativistic and chemical enhancements of $\mathcal{P,T}$-odd effects in polar diatomic radicals\footnote{Parts of this work were
created during the master thesis of Konstantin Gaul.}}
\date{\today}
\author{Konstantin Gaul}
\affiliation{Fachbereich Chemie, Philipps-Universit\"{a}t Marburg,
Hans-Meerwein-Stra\ss{}e 4, 35032 Marburg, Germany}
\author{Sebastian Marquardt}
\affiliation{Fachbereich Chemie, Philipps-Universit\"{a}t Marburg,
Hans-Meerwein-Stra\ss{}e 4, 35032 Marburg, Germany}
\author{Timur Isaev}
\affiliation{Petersburg Nuclear Physics Institute, Orlova Roscha. 1,
188300 Gatchina, Russia}
\author{Robert Berger}
\affiliation{Fachbereich Chemie, Philipps-Universit\"{a}t Marburg,
Hans-Meerwein-Stra\ss{}e 4, 35032 Marburg, Germany}
\begin{abstract}

Polar diatomic molecules that have, or are expected to have a $^2\Sigma_{1/2}$-ground state are
studied systematically with respect to simultaneous violation of parity
$\mathcal{P}$ and time-reversal $\mathcal{T}$ with numerical
methods and analytical models. Enhancements of
$\mathcal{P,T}$-violating effects due to an electric dipole moment of the electron (eEDM)
and $\mathcal{P,T}$-odd scalar-pseudoscalar nucleon-electron current
interactions are
analyzed by comparing trends within columns and rows of the periodic
table of the elements. For this purpose electronic structure
parameters are calculated numerically within a quasi-relativistic zeroth order regular
approximation (ZORA) approach in the framework of complex generalized
Hartree-Fock (cGHF) or Kohn-Sham (cGKS). Scaling
relations known from analytic relativistic atomic structure
theory are compared to these numerical results. Based on this analysis, problems of commonly used relativistic enhancement
factors are discussed. Furthermore the
ratio between both $\mathcal{P,T}$-odd electronic structure
parameters mentioned above is analyzed for various groups of the
periodic table. From this analysis an analytic measure for the
disentanglement of the two $\mathcal{P,T}$-odd electronic structure
parameters with multiple experiments in dependence of electronic
structure enhancement factors is derived.  
\end{abstract}
\maketitle
\section{Introduction}
Simultaneous violation of space- ($\mathcal{P}$) and time-parity
($\mathcal{T}$) in the charged lepton sector is considered to be a strong indicator for physics beyond the standard
model of particle physics\cite{gross:1996}. 
Exploiting enhancement effects in bound systems, such as atoms or
molecules, low-energy experiments actually provide the best limits on
$\mathcal{P,T}$-violation and thus are among the most useful tools to exclude
new physical theories and to test the Standard
Model\cite{fortson:2003,khriplovich:1997}.\par
Understanding  these atomic and molecular enhancement effects in detail
is essential for the development of sensitive experiments.\par
A permanent atomic or molecular
electric dipole moment (EDM) that causes a linear Stark shift in the limit
of zero external fields would violate
$\mathcal{P,T}$.\cite{khriplovich:1997}
Mainly four sources of a permanent EDM in
molecules are considered: permanent electric dipole moments of the nuclei,
$\mathcal{P,T}$-odd nucleon-nucleon current interactions, a
permanent electric dipole moment of the
electron (eEDM) and $\mathcal{P,T}$-odd nucleon-electron current interactions (see e.g.
\cite{ginges:2004}). Of these sources the latter two have the most
important contribution in paramagnetic systems.\cite{ginges:2004}
Furthermore, nucleon-electron
interactions are expected to be dominated by scalar-pseudoscalar
interactions, that
are nuclear spin independent.\par
Since the formulation of an eEDM interaction Hamiltonian for atoms by
Salpeter in the year 1958\cite{salpeter:1958},
there have been many studies on eEDM enhancement in atoms and
molecules. Sandars worked out analytical relations of atomic eEDM
interactions in the 1960s
\cite{sandars:1965,sandars:1966,sandars:1968,sandars:1968a}, which
where confirmed also by others some time
later\cite{ignatovich:1969,flambaum:1976}. Sandars calculated, that
the enhancement of the eEDM in atoms scales with $\alpha^2Z^3$, where
$\alpha$ is the fine-structure constant and $Z$ is the nuclear charge number.
Enhancements of scalar-pseudoscalar
nucleon-electron current interactions in atoms scale
as $\alpha Z^3$, as well\cite{khriplovich}.
Since then, a number of numerical studies was conducted, but most of the 
previous
investigations focused on the description of $\mathcal{P,T}$-odd effects in
individual or few molecular candidates.\par
Some attempts were made to obtain a deeper
understanding of enhancement of $\mathcal{P,T}$-odd effects in
molecules beyond established $Z$-dependent scaling laws. In 
Ref.~\onlinecite{prasannaa:2015}, for instance, 
the influence of the nuclear charge number of the
electronegative partner on eEDM enhancements in mercury monohalides
was studied. Furthermore  effects of the
polarization of the molecule by the electronegative partner on the eEDM
enhancement are discussed. In Ref.~\onlinecite{prasannaa:2015} it was
concluded that the nuclear charge of the lighter halogen atom has lower
influence on the eEDM enhancement than its electronegativity.\par
Recently Sunaga \text{et. al.} studied large eEDM enhancement effects in hydrides within orbital
interaction theory and remarked an influence of the energy difference
between the interacting valence orbitals of the electronegative atom
and the unoccupied p$_{1/2}$ orbital of the heavy
atom\cite{sunaga:2017}. Both of the mentioned studies confirmed that large contributions of s- and p-type
atomic orbitals in the singly occupied molecular orbital increase
$\mathcal{P,T}$-odd effects, as expected\cite{khriplovich}. A
similar result was obtained by Ravaine
 \text{et. al.} in 2005\cite{revaine:2005}, who showed that the covalent
character of HI$^+$ causes a stronger s-p-mixing and therefore a larger
enhancement of the eEDM than in HBr$^+$,
which has an ionic bond.\par
The majority of previous studies on $\mathcal{P,T}$-violating effects in
molecules were performed within a four-component (relativistic)
framework. Our recently
developed two-component (quasi-relativistic) approach for the calculation of
$\mathcal{P,T}$-odd effects allows for routine calculations of
a large number of molecules on an ab initio level (see
Ref.~\onlinecite{gaul:2017} for details of the method). In our present paper
we thus study systematically a
wealth of diatomic
radicals across the periodic table, which are known to have a
$^2\Sigma_{1/2}$-ground state, or for which at least a
$^2\Sigma_{1/2}$-ground state can naively be expected from simple
chemical bonding concepts. In combination with
analytic scaling relations we calculate the $Z$-dependent
and $Z$-independent electronic structure effects in different groups
of the periodic table. Furthermore with an analysis of the behavior of
isolobal diatomic molecules we  gauge in particular the "chemical"
influences on the $\mathcal{P,T}$-odd enhancement, that is new
effects that change between different columns of the periodic table.\par 
We provide with our analysis a consistent overview of $\mathcal{P,T}$-odd effects
in a large number of diatomic molecules, which serves as a suitable starting 
point for further research with higher-level electronic structure methods, 
where needed. By
analysing general trends of the ratio between molecular enhancement factors 
of the electron electric dipole moment and nucleon-electron current 
interactions, we draw conclusions on possibilities to disentangle 
them in experiments with polar 
diatomic radicals that feature a $^2\Sigma_{1/2}$-ground state.

\section{Theory\label{theory}}
\subsection{$\mathcal{P,T}$-odd spin-rotational Hamiltonian
\label{spinrot}}
We present herein electronic structure calculations for polar diatomic molecules
that are expected to have a $^2\Sigma_{1/2}$-ground state. For these systems an
effective spin-rotational Hamiltonian can be derived that in
particular describes a
transition of Hund's coupling case (c)  to case
(b)\cite{hund:1927a,hund:1927b,hund:1927}. This corresponds to cases,
where the rotational constant is much smaller than the spin-doubling
constant but much larger than the $\Omega$-doubling constant (for
details see Ref.~\onlinecite{kozlov:1995}). The $\mathcal{P,T}$-odd part of this effective spin-rotational
Hamiltonian reads (see e.g. Refs.~\onlinecite{dmitriev:1992,kozlov:1995})
\begin{equation}
H_\text{sr}=\parantheses{k_\text{s}W_\text{s}+d_\text{e}W_\text{d}}\Omega
           =W_\text{d} \parantheses{k_\text{s}W_\text{s}/W_\text{d}+d_\text{e}}\Omega,
\label{eq: spinrot}
\end{equation}
where $\Omega=\vec{J}_\text{e}\cdot\vec{\lambda}$ is the projection of
the reduced total electronic angular momentum $\vec{J}_\text{e}$ on the
molecular axis, defined by the unit vector $\vec{\lambda}$ pointing
from the heavy to the light nucleus. $k_\text{s}$ is the
$\mathcal{P,T}$-odd
scalar-pseudoscalar nucleon-electron current interaction constant and
$d_\text{e}$ is the eEDM. The $\mathcal{P,T}$-odd
electronic structure parameters are defined by
\begin{subequations}
\begin{align}
W_\text{s}&=\frac{\Braket{\Psi|\Op{H}_\text{s}|\Psi}}{k_\text{s}\Omega}\\
W_\text{d}&=\frac{\Braket{\Psi|\Op{H}_\text{d}|\Psi}}{d_\text{e}\Omega},
\end{align}
\end{subequations}
where $\Psi$ is the electronic wave function and the molecular
$\mathcal{P,T}$-odd Hamiltonians are\cite{salpeter:1958,khriplovich:1997}:
\begin{align}
\Op{H}_\text{s}&=\i
k_{\text{s}}\frac{G_\text{F}}{\sqrt{2}}\Sum{i=1}{N_\text{elec}}
\Sum{A=1}{N_\text{nuc}}\rho_A\parantheses{\pos_i}Z_A\diraccontra{0}\diraccontra{5}
\label{eq: eNcpviol}\\
\Op{H}_{\text{d}}&=-d_\text{e}\Sum{i=1}{N_\text{elec}}\parantheses{\diraccontra{0}-1}
\vec{\boldsymbol{\Sigma}}\cdot\vec{\efield}(\vec{r}_i).
\label{eq: stratagemI}
\end{align}
Here $\rho_A$ is the normalized nuclear density distribution of nucleus
$A$ with charge number $Z_A$, $\pos_i$ is the position vector of
electron $i$, $\vec{\efield}$ is the internal electrical field,
$G_\text{F}=2.22249\times10^{-14}~E_\text{h}a_0^3$ is Fermi's weak
coupling constant, $\i=\sqrt{-1}$ is the imaginary unit and the Dirac
matrices in
standard notation are defined as ($k=1,2,3$)
\begin{equation}
\begin{gathered}
\diraccontra{0}=\begin{pmatrix}
\unity & \zeroty\\
\zeroty & -\unity 
\end{pmatrix},~~
\diraccontra{k}=\begin{pmatrix}
\zeroty&\pauli^k\\
-\pauli^k&\zeroty
\end{pmatrix},\\
\diraccontra{5}=
\begin{pmatrix}
\zeroty&\unity\\
\unity&\zeroty
\end{pmatrix},~~
\boldsymbol{\Sigma}^k=
\begin{pmatrix}
\pauli^k&\zeroty\\
\zeroty&\pauli^k
\end{pmatrix}
\end{gathered}
\end{equation}
with the vector of the Pauli spin matrices $\vec{\pauli}$. 
$\Op{H}_\text{d}$ as reported here is obtained according to Stratagem
I of Ref.~\onlinecite{lindroth:1989} by commuting the unperturbed
Dirac-Coulomb Hamiltonian with a modified momentum operator. We will
come back to this in Section \ref{manyel}.

In this work the electronic structure parameters were calculated,
using the corresponding quasi-relativistic Hamiltonians within the
zeroth order regular approximation
(ZORA)\cite{Isaev:2013,kudashov:2014,gaul:2017}
\begin{align}
\Op{H}^\text{ZORA}_{\text{s}}=&\i\Sum{i=1}{N_\text{elec}}
\Sum{A=1}{N_\text{nuc}}Z_A\brackets{\rho_A(\pos_i)
\tilde{\omega}_\text{s}(\pos_i),\spinmom_i}_-.
\label{eq: zoraeNPT},\\
\Op{H}^\text{ZORA}_{\text{d}}
=&\Sum{i=1}{N_\text{elec}}\parantheses{\spinmom_i}\tilde{\omega}_{\text{d}}(\pos_i)
\vec{\Pauli}\cdot\vec{\efield}(\pos_i)\parantheses{\spinmom_i}
\label{eq: ZORA},
\end{align}
where $\momop$ is the linear momentum operator,
$\brackets{A,B}_-=AB-BA$ is the commutator and
the modified ZORA factors are defined as
\begin{align}
\tilde{\omega}_\text{s}(\pos_i)&=\frac{G_\text{F}k_\text{s}c}{\sqrt{2}\parantheses{2m_\text{e}c^2-\tilde{V}(\pos_i)}}~~,\\
\tilde{\omega}_\text{d}(\pos_i)&=\frac{2d_\text{e}c^2}{\parantheses{2m_\text{e}c^2-\tilde{V}(\pos_i)}^2}~~,
\end{align}
with the model potential $\tilde{V}$ introduced by van
W\"ullen\cite{wullen:1998},
which is used to alleviate the gauge dependence of ZORA. Here $c$ is
the speed of light in vacuum and
$m_\text{e}$ is the mass of
the electron. The internal electrical field can be approximated as the
field of the nuclei\cite{lindroth:1989,gaul:2017}:
\begin{equation}
\vec{\efield}(\pos_i)\approx\Sum{A=1}{N_\text{nuc}}k_\text{es}Z_A
e\frac{\pos_i-\pos_A}{\abs{\pos_i-\pos_A}^3},
\label{eq: efieldapprox}
\end{equation}
with $e$ being the elementary
charge and the constant $k_\text{es}$ being
$\parantheses{4\pi\epsilon_0}^{-1}$ in SI units with the electric
constant $\epsilon_0$. Furthermore the total angular
momentum projection was calculated explicitly by
\begin{multline}
\Omega=\left(\Braket{\Psi_\text{ZORA}|\Sum{i}{}\hat{\vec{\ell}}_i|\Psi_\text{ZORA}}+
\right.\\\left.
\frac{1}{2}\Braket{\Psi_\text{ZORA}|\Sum{i}{}\vec{\pauli}_i|\Psi_\text{ZORA}}\right)\cdot \vec{\lambda},
\end{multline}
where $\hat{\vec{\ell}}_i$ is the reduced orbital angular momentum
operator for electron $i$ and $\Psi_\text{ZORA}$ is the ZORA multi-electron
wave function.

\subsection{Scaling-relations of $\mathcal{P,T}$-odd properties}
Within the relativistic Fermi-Segr\`e model for electronic wave
functions\cite{fermi:1933} the matrix elements of the $\mathcal{P,T}$-odd operators can be obtained
analytically for atomic systems\cite{bouchiat:1974,khriplovich}.  
The results for the $\mathcal{P,T}$-odd nucleon-electron current
interactions can be expressed in
terms of a relativistic enhancement
factor
\begin{equation}
R(Z,A)=\frac{4}{\Gamma^2\parantheses{2\gamma+1}}\parantheses{2Zr_\text{nuc}/a_0}^{2\gamma-2},
\end{equation}  
where $\Gamma(z)$ is the gamma function, $Z$ and $A$ are the nuclear
charge and mass numbers, respectively, $r_\text{nuc}\approx1.2~\text{fm}\cdot
A^{1/3}$ is the nuclear radius, $a_0$ is the Bohr radius and 
\begin{equation}
\gamma=\sqrt{\parantheses{j+\frac{1}{2}}^2-\parantheses{\alpha Z}^2},
\end{equation}
with the fine structure constant $\alpha\approx\frac{1}{137}$ and the
total electronic angular momentum quantum number $j$.\par
In terms of the relativistic enhancement the parameters of the
$\mathcal{P,T}$-odd spin-rotational Hamiltonian can now be estimated to behave as
(see Ref.~\onlinecite{khriplovich} for  $W_\text{s}$ and Ref.~
\onlinecite{flambaum:1976} and Ref.~\onlinecite{sushkov:1978} for
$W_\text{d}$) 
\begin{align}
W_\text{s}&\approx -\frac{G_\text{F}}{2\pi\sqrt{2}a_0^3}\underbrace{R(Z,A)\gamma}_{R_\text{s}(Z,A)}
Z^3\alpha \varkappa\label{eq: wsscale},\\
W_\text{d}&\approx-\frac{4 E_\text{h}}{3e\cdot a_0}
\underbrace{\frac{3}{\gamma\parantheses{4\gamma^2-1}}}_{R_\text{d,CS}(Z)}
Z^3\alpha^2 \varkappa\label{eq: wdscale},
\end{align}
where $\varkappa$ is a constant that depends on the effective 
electronic structure of the system under study.\par
\begin{figure}\centering
\resizebox{.5\textwidth}{!}{\includegraphics{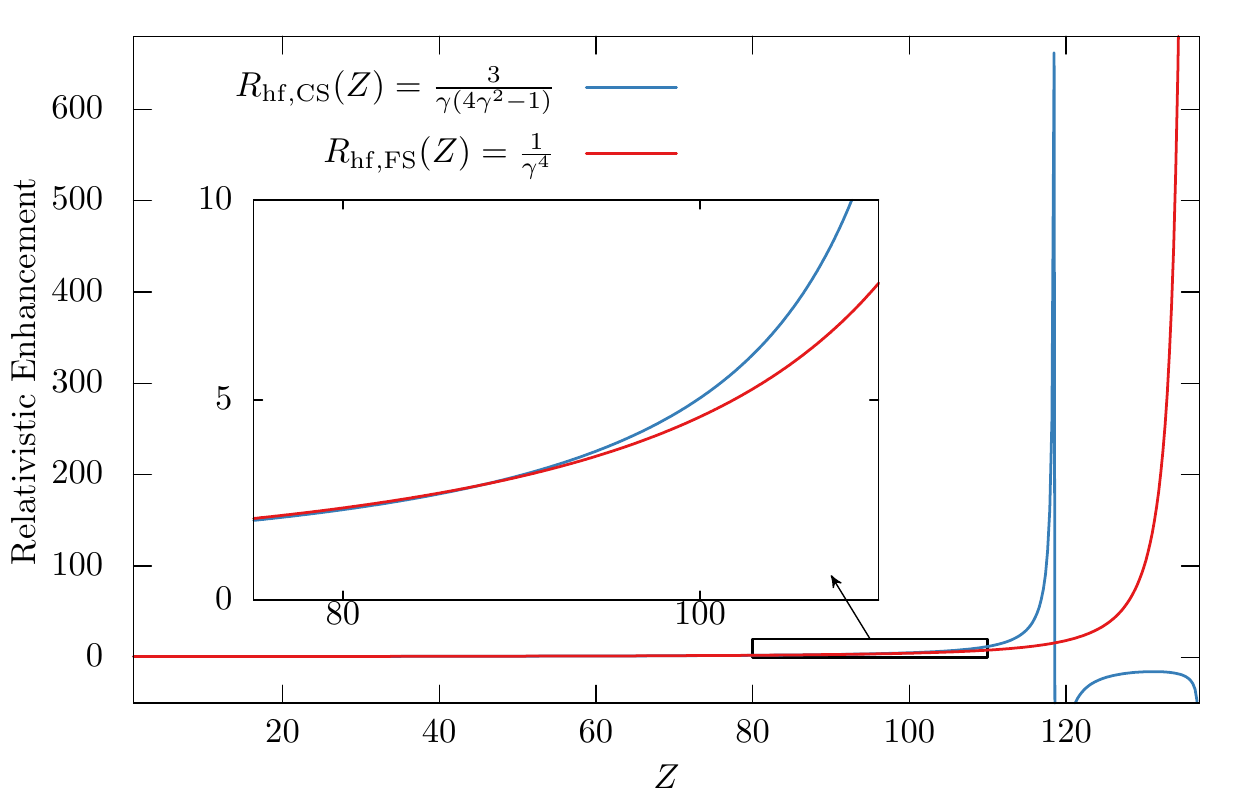}}
\caption{
Comparison of relativistic enhancement factors for eEDM induced
permanent EDMs of atoms. Factor by Sandars derived analytically with Casimirs method
(CS) and empirical factor for hyperfine interaction found by Fermi and
Segr\`e (FS). Plots are shown for the case of $j=\frac{1}{2}$ as in
$^2\Sigma_{1/2}$-states.
}
\label{fig: relfacs}
\end{figure}

We note in passing, that the relativistic enhancement factor 
of the eEDM induced permanent atomic EDM $R_\text{d,CS}(Z)$ is the
same as the one for hyperfine interactions published first by Racah in
1931\cite{racah:1931}. In relation \prettyref{eq: wdscale} the label CS indicates that the factor was
derived by Sandars\cite{sandars:1966} from a method by Casimir. The
denominator in relation \prettyref{eq: wdscale} has two roots: one at $Z=\frac{\sqrt{j^2+j}}{\alpha}$ and
one at $Z=\frac{\frac{1}{2}+j}{\alpha}$.
Thus the relativistic enhancement factor causes problems not only for
$Z>137$ but diverges at
$Z=\frac{\sqrt{3}}{2\alpha}\approx118.65$ for $^2\Sigma_{1/2}$-states
(see \prettyref{fig: relfacs}).
This was also found by Dinh \textit{et. al.} in a study of hyperfine
interactions in super heavy atoms.\cite{dinh:2009} 
These findings imply that relation \prettyref{eq: wdscale} is of
limited use to estimate $W_\text{d}$ for elements with $Z>100$.\par
An alternative relativistic enhancement factor for
hyperfine interactions was found empirically by Fermi and Segr\`e
\cite{fermi:1933a,fermi:1933}, who interpolated numerically calculated data by Racah and
Breit\cite{racah:1931,breit:1931}:
\begin{equation}
R_\text{hf,FS}(Z)= \frac{1}{\gamma^4}\label{eq: wdscaleemp},
\end{equation}
where the label FS was introduced referring to Fermi and Segr\`e.
$R_\text{hf,FS}(Z)$ has no singularities for $Z<137$, and therefore no severe
problems in the description of elements up to $Z\leq118$ are
expected. Furthermore, eq. \prettyref{eq: wdscaleemp} can also be applied to
estimate the eEDM enhancement, because the atomic integrals relevant
for the hyperfine structure and eEDM enhancement do not differ much within the
Fermi-Segr\`e model and result in the similar enhancement
factors differing only by a factor of $\alpha Z$ (see also above and \cite{sandars:1966}):
\begin{subequations}
\begin{align}
R_\text{hf}(Z)\sim&\int\text{d}r~r^{-2}g_0(r)f_0(r),\\
R_\text{d}(Z)\sim&\int\text{d}r~r^{-2}f_1(r)f_0(r),
\end{align}
\end{subequations}
where $g_\ell$ and $f_\ell$ are the upper and lower component of the Dirac
bi-spinor for a specific orbital angular quantum number $\ell$,
respectively. As $g_0(r)\alpha Z\approx
f_1(r)+\text{corrections}$, for hydrogen-like atoms, the
relativistic enhancement factors are in a first approximation
identical up to a factor of $\alpha Z$. Thus the empirical factor \prettyref{eq: wdscaleemp} can
be
employed for our purposes (see also \prettyref{fig: relfacs}).\par 
An improved relativistic enhancement factor for the
$\mathcal{P,T}$-odd nucleon-electron current interaction parameter $W_\text{s}$ was 
calculated with an analytical atomic model in \cite{dzuba:2011}:
\begin{equation}
W_\text{s}\approx \frac{G_\text{F}}{2\pi\sqrt{2}a_0^3} Z^3\alpha
R(Z,A)f(Z) \frac{\gamma+1}{2}\varkappa
\label{eq: dzubaws} 
\end{equation}
with the $Z$-dependent function
\begin{equation}
f(Z)=\frac{1-0.56\alpha^2Z^2}{\parantheses{1-0.283\alpha^2Z^2}^2},
\label{eq: scalefunc}
\end{equation}
which results from a polynomial expansion of the atomic wave functions
(see appendix of Ref.~\onlinecite{dzuba:2011} for details\footnote{The
explicit numerical factors in $f(Z)$ were printed partially
wrong in Ref.~\onlinecite{dzuba:2011}, which was
mentioned in Ref.~\onlinecite{Isaev:2013}.}). In Refs.
\onlinecite{dzuba:2011,Isaev:2013} the eEDM enhancement parameter
$W_\text{d}$ was estimated from $W_\text{s}$ by use of a relativistic 
enhancement factor for the ratio
$W_\text{d}/W_\text{s}$ derived from eqs. \prettyref{eq: dzubaws} and
\prettyref{eq: wdscale}:  
\begin{equation}
\tilde{R}_\text{CS}(Z,A)=\frac{6}%
{\gamma\parantheses{4\gamma^2-1}\parantheses{\gamma+1}\cdot f(Z) R(Z,A)}.
\label{eq: wdfromwscs}
\end{equation}
In combination with summarized conversion factors and constant
pre-factors of $W_\text{s}$ and $W_\text{d}$
\begin{equation}
c_\text{conv} =
\frac{8\sqrt{2}\pi\alpha}{3\frac{G_\text{F}\cdot e}{E_\text{h}a_0^2}},
\end{equation}
where $G_\text{F}$ is Fermi's constant in atomic units, an estimate for
$W_\text{d}$ is received from $W_\text{s}$ via
\begin{equation}
W_\text{d}\approx c_\text{conv}\cdot \tilde{R}_\text{CS}(Z,A) W_\text{s}.
\end{equation} 
When relation \prettyref{eq: wdscaleemp} is
used instead of \prettyref{eq: wdscale}, one obtains an alternative
relativistic enhancement factor, which is expected to be more accurate
for atoms with a high $Z$:
\begin{equation}
\tilde{R}_\text{FS}(Z,A)=\frac{2}{\gamma^4\parantheses{\gamma+1}\cdot R(Z,A) f(Z)}.
\label{eq: wdfromwsfs}
\end{equation}
For comparison, instead of the improved relativistic factor for
$W_\text{s}$ (eq. \prettyref{eq: dzubaws}) relation \prettyref{eq: wsscale}
can be used to receive relativistic enhancement factors:
\begin{subequations}
\begin{align}
\tilde{\tilde{R}}_\text{CS}(Z,A)&=\frac{3}%
{\gamma^2\parantheses{4\gamma^2-1}\cdot R(Z,A)},\label{eq: wdwscs}
\\
\tilde{\tilde{R}}_\text{FS}(Z,A)&=\frac{1}{\gamma^5\cdot
R(Z,A)}.\label{eq: wdwsfs}
\end{align}
\end{subequations}
In the following discussion we will show that eq. \prettyref{eq:
wdscaleemp} and \prettyref{eq: wdfromwsfs} indeed agree much better
with numerical calculations for $Z>100$ than eq. \prettyref{eq:
wdscale} and \prettyref{eq: wdfromwscs}, while there is no appreciable
difference for molecules with lighter atoms.

\subsection{Neglected many-electron effects in light molecules\label{manyel}}
The $\mathcal{P,T}$-odd operators shown in \ref{spinrot} are
one-electron operators. Their expectation values scale with the
nuclear charge number as $Z^3$. Thus these contributions are
dominant in high-$Z$
molecules. However, in light molecules many-electron
effects with lower $Z$-dependence stemming from the Hartree--Fock
picture or the Breit interaction  can have an 
important contribution to the enhancement factors. \par
In the following we focus first on additional contributions in
the Dirac--Hartree--Fock (DHF) picture that arise from the ZORA
transformation. The DHF equation without magnetic fields and with perturbations
\prettyref{eq: stratagemI} and
\prettyref{eq: eNcpviol} reads
\begin{widetext}
\begin{equation}
\begin{pmatrix}
\Op{V}_0(\pos_i)\unity-\Op{\bf K}_{\phi\phi}-\epsilon_i\unity
&c\spinmom_i-\Op{\bf K}_{\phi\chi}+\i
k_\text{s}\frac{G_\text{F}}{\sqrt{2}}\rho_\text{nuc}(\pos_i)\unity\\
c\spinmom_i-\Op{\bf K}_{\chi\phi}-\i
k_\text{s}\frac{G_\text{F}}{\sqrt{2}}\rho_\text{nuc}(\pos_i)\unity&\parantheses{\Op{V}_0(\pos_i)-2m_\text{e}c^2}\unity-\Op{\bf K}_{\chi\chi}-\epsilon_i\unity+2d_\text{e}\vec{\pauli}\cdot\vec{\efield}(\pos_i)
\end{pmatrix}
\begin{pmatrix}
\phi_i\\
\chi_i
\end{pmatrix}
=
\begin{pmatrix}
0\\
0
\end{pmatrix}
\label{eq: dhfeq}
\end{equation}
\end{widetext}
where $\phi_i$ and $\chi_i$ are the upper and lower components of the
Dirac bi-spinor of electron $i$, respectively and $\epsilon_i$ is its
orbital energy. The nuclear charge density is summarized as
$\rho_\text{nuc}(\pos_i)=\Sum{A=1}{N_\text{nuc}}Z_A\rho_A(\pos_i)$
and 
$\Op{V}_0(\pos_i)=\Op{V}_\text{ext}(\pos_i)+\Op{V}_\text{nuc}(\pos_i)+\Op{J}_{\phi\phi}(\pos_i)+\Op{J}_{\chi\chi}(\pos_i)$
is the potential energy operator appearing on the diagonal, where $\Op{V}_\text{ext}$ and $\Op{V}_\text{nuc}$ are the external and
nuclear potential energy operators, respectively. $\Op{J}_{\phi\phi}$
and $\Op{J}_{\chi\chi}$ are the direct parts and $\Op{\bf K}_{\phi\phi}$,
$\Op{\bf K}_{\phi\chi}$, $\Op{\bf K}_{\chi\phi}$, $\Op{\bf K}_{\chi\chi}$ are the
exchange parts that emerge from the two-electron Coulomb operator in
DHF theory. From here on we drop the electron index and the
dependencies on the electronic positions for better readability.\par
Whereas the direct Dirac-Coulomb contributions $\Op{J}_{\phi\phi}$ and
$\Op{J}_{\chi\chi}$ are local and appear on
the diagonal, the exchange contributions are non-local and non-diagonal 
\begin{equation}
\Op{\bf K}=\begin{pmatrix}
\Op{\bf K}_{\phi\phi}&&\Op{\bf K}_{\phi\chi}\\
\Op{\bf K}_{\chi\phi}&&\Op{\bf K}_{\chi\chi}
\end{pmatrix}.
\end{equation}
Thus when deriving an approximate relation between $\phi$ and $\chi$,
as when transforming into the ZORA picture, the exchange terms can result
in additional contributions to the $\mathcal{P,T}$-odd
enhancement.\par
We start our discussion with the scalar-pseudoscalar nucleon-electron current interaction
Hamiltonian. The ZORA Hamiltonian within this perturbation appears as
\begin{multline}
\Op{H}_0^\text{ZORA-HF}+
\Op{H}_\text{s}^\text{ZORA-HF}=\\
\parantheses{\spinmom
-\frac{1}{c}\Op{\bf K}_{\phi\chi}+\i
k_\text{s}\frac{G_\text{F}}{c\sqrt{2}}\rho_\text{nuc}\unity}\omega\\
\times
\parantheses{\spinmom-\frac{1}{c}\Op{\bf K}_{\chi\phi}-\i
k_\text{s}\frac{G_\text{F}}{c\sqrt{2}}\rho_\text{nuc}\unity},
\end{multline}
where
$\Op{H}_0^\text{ZORA-HF}=\parantheses{\spinmom-\frac{1}{c}\Op{\bf
K}_{\phi\chi}}\omega\parantheses{\spinmom-\frac{1}{c}\Op{\bf
K}_{\chi\phi}}$ is the unperturbed ZORA Hamiltonian in
the HF approximation and $\omega=\frac{c^2}{2m_\text{e}c^2-\tilde{V}}$ is the
ZORA-factor with the model potential $\tilde{V}$.
This results in additional correction terms to \prettyref{eq:
zoraeNPT} stemming from the many-electron mean-field picture (only terms to first
order in $G_\text{F}$ are shown):
\begin{equation}
\Delta\Op{H}_\text{s}^\text{ZORA-HF}=
\frac{1}{c}\rho_\text{nuc}\tilde{\omega}_\text{s}\Op{\bf K}_{\chi\phi}
-\frac{1}{c}\Op{\bf K}_{\phi\chi}\tilde{\omega}_\text{s}\rho_\text{nuc}
\label{eq: zorahfcorreNc}
\end{equation}
As $\tilde{\omega}_\text{s}$ and
the exchange operators $\Op{\bf K}_{\phi\chi}$,$\Op{\bf K}_{\chi\phi}$
are of $\mathcal{O}(\alpha)$, that is of the order of $\alpha$, these
corrections are of $\mathcal{O}(\alpha^3)$, whereas 
the Hamiltonian defined in eq. \prettyref{eq: eNcpviol} is of first order in $\alpha$.\par 
We now focus on the eEDM interaction Hamiltonian. The ZORA transformation of the DHF operator using our method from \cite{gaul:2017} yields:
\begin{equation}\Op{H}_\text{d}^\text{ZORA-HF}=
\parantheses{\spinmom-\frac{1}{c}\Op{\bf K}_{\phi\chi}}
\parantheses{\tilde{\omega}_\text{d}\vec{\pauli}\cdot\vec{\efield}}
\parantheses{\spinmom-\frac{1}{c}\Op{\bf K}_{\chi\phi}}
\end{equation}
Thus many-electron mean-field correction terms to \prettyref{eq: ZORA}
are received as
\begin{multline}
\Delta\Op{H}_\text{d}^\text{ZORA-HF}=
-\frac{1}{c}\Op{\bf K}_{\phi\chi}\tilde{\omega}_\text{d}\vec{\pauli}\cdot\vec{\efield}\spinmom\\
-\frac{1}{c}\spinmom\tilde{\omega}_\text{d}\vec{\pauli}\cdot\vec{\efield}\Op{\bf K}_{\chi\phi}
+\frac{1}{c^2}\Op{\bf K}_{\phi\chi}\tilde{\omega}_\text{d}\vec{\pauli}\cdot\vec{\efield}\Op{\bf K}_{\chi\phi}.
\label{eq: zorahfcorr}
\end{multline}
The terms are sorted by their order in the fine structure constant
$\alpha$. The first two terms
are of $\mathcal{O}(\alpha^4)$ and the last term is of 
$\mathcal{O}(\alpha^6)$ and thus is suppressed. The first two terms are suppressed by a factor
$\alpha^2$ in comparison to the operator of eq. \prettyref{eq:
ZORA}. This is why the correction terms of eq. \prettyref{eq:
zorahfcorreNc} and \prettyref{eq: zorahfcorr} have been neglected in
the present study even when HF is used. For light elements, however,
such terms can be more important, as has been shown e.g. in Ref.
\cite{berger:2008} \par 
In a density functional theory (DFT) picture none of the above terms
$\Delta\Op{H}_\text{d}^\text{ZORA-HF}$,
$\Delta\Op{H}_\text{s}^\text{ZORA-HF}$ arises if conventional
non-relativistic density functionals are used. Thus we would expect a
larger deviation of HF-ZORA from DHF calculations than of Kohn--Sham
(KS)-ZORA from
Dirac--Kohn--Sham (DKS) calculations. However, if hybrid functionals are
used as in our present paper, Fock-exchange is considered explicitly
and inclusion of the correction terms mentioned above may become
necessary for light elements. \par 
If the above discussed exchange terms become important, terms of
comparatively low order which
are so far neglected may become important, too. These include the two-electron part of the internal
electrical field 
\begin{equation}
-\Sum{i<j}{N_\text{elec}}k_\text{es}
e\parantheses{\diraccontra{0}-1}\vec{\boldsymbol{\Sigma}}\cdot\frac{\pos_i-\pos_j}{\abs{\pos_i-\pos_j}^3}.
\end{equation}
However, if an alternative effective one-electron form of operator \prettyref{eq:
stratagemI} is used, the two-electron contributions from the electric
field can be included implicitly within a mean-field approach.\cite{martensson-pendrill:1987} Our
previous calculations\cite{gaul:2017} have shown, that these effects are negligible
and even for very light molecules as boron monoxide the effects are
below 5~\%{}
(see Supplemental Material)
and are thus not important for the present discussion.\par 
Another term of comparatively low order in $\alpha$ is the Breit contribution.
The transformed form of the Breit contributions to
eEDM enhancement, that corresponds to \prettyref{eq: stratagemI} was
derived in \cite{lindroth:1989}:
\begin{multline}
\Op{H}^\text{Breit}_\text{d}=\frac{d_\text{e}e}{\hbar}\Sum{i\not=j}{}\i\brackets{\diraca_i\cdot\momop_i,
\right.\\\left.
\parantheses{\frac{k_\text{es}}{2}\frac{\vec{\boldsymbol{\Sigma}}_i\cdot\diraca_j+\vec{\boldsymbol{\Sigma}}_i\cdot\frac{\pos_i-\pos_j}{\abs{\pos_i-\pos_j}}\diraca_j\cdot\frac{\pos_i-\pos_j}{\abs{\pos_i-\pos_j}}}{\abs{\pos_i-\pos_j}}}}.
\end{multline}
Here we introduced the Dirac matrix
$\diraca=\diraccontra{0}\vec{\diraccontra{}}$.
Additional corrections appear from the ZORA
transformation, when the Breit interaction, which appears as well on the 
off-diagonal, is considered (see e.g.
\cite{berger:2008}). These Breit interaction corrections  appear for $\Op{H}_\text{s}$
as well.\par
For a more accurate calculation of the eEDM enhancement other magnetic terms of
$\mathcal{O}(\alpha^2)$, which were neglected in the deviation in our previous
paper\cite{gaul:2017}, can play an important role as well and should be
considered (see e.g. \cite{lindroth:1989}). For the operator used in
this work (eq. \prettyref{eq: stratagemI}) these are

\begin{equation}
\Op{H}^\text{mag}_\text{d}=\i
d_\text{e}\brackets{\parantheses{\diraccontra{0}-1}\diraca\cdot\vec{\bfield}-\diraccontra{5}\parantheses{\vec{A}\cdot\momop+\momop\cdot\vec{A}}}
\end{equation}
and choosing Coulomb gauge within ZORA they appear as
\begin{multline}
\Op{H}^\text{mag,ZORA}_\text{d}=-\i d_\text{e}
\parantheses{
2\spinmompot\omega\vec{\pauli}\cdot\vec{\bfield}
\right.\\\left.
-\spinmompot\omega\unity\vec{A}\cdot\nabla -
\unity\vec{A}\cdot\nabla\omega\spinmompot
}
\end{multline}
where $\hat{\vec{\pi}}=\momop+e\vec{A}$ with the vector potential
$\vec{A}$.
Additional magnetic contributions arise from the ZORA
transformation due to the vector potential on the off-diagonal.
\par
Regarding many-body
effects of the operator itself, things would become more
complicated in a DFT picture, where only one-electron
operators are well-defined. Whereas the direct contribution could be calculated
analogously to HF, an correction term to the exchange-correlation
potential would appear and special exchange-correlation energy
functionals would have to be designed. In case of hybrid DFT,
additionally Fock exchange contributions would have to be
computed. Herein, however, an inclusion of such
correction terms is not attempted.\par
In our present calculations all these many-electron operators are neglected. 
In principle, this could cause a deviation from
comparable four-component calculations which becomes in relative terms more 
pronounced in light molecules 
than in high-$Z$ molecules and are expected to mainly originate from the terms \prettyref{eq: zorahfcorreNc} and \prettyref{eq: zorahfcorr}. 
But these are still expected to be small.\par

\section{Computational Details}
Quasi-relativistic two-component calculations are performed within 
ZORA at the level of complex generalized Hartree--Fock (cGHF) or
Kohn--Sham (cGKS) with a modified 
version\cite{berger:2005a,berger:2005,isaev:2012,nahrwold:09,gaul:2017} of the
quantum chemistry program package Turbomole\cite{ahlrichs:1989}. 
In order to calculate the $\mathcal{P,T}$-odd properties, the
program was extended with the corresponding ZORA Hamiltonians (see
\cite{gaul:2017} for details on the implementation).\par
For Kohn--Sham (KS)-density functional theory (DFT) calculations the hybrid Becke three parameter exchange
functional and Lee, Yang and Parr correlation functional
(B3LYP)\cite{stephens:1994,vosko:1980,becke:1988,lee:1988} was employed.
For all calculations a basis set of 37~s, 34~p, 14~d and 9~f
uncontracted Gaussian functions with the exponential coefficients
$\alpha_i$ composed as an even-tempered series as $\alpha_i=a\cdot
b^{N-i};~ i=1,\dots,N$, with $b=2$ for s- and p-function and with
$b=(5/2)^{1/25}\times10^{2/5}\approx 2.6$ for d- and f-functions
was used for the electro-positive atom (for details see Supplementary Material).
\footnote{For the calculation of row 8 compounds the basis set was augmented
with more diffuse functions and a set of g-functions. However, these
showed no remarkable influence on $\mathcal{P,T}$-odd properties and
thus the results for the same basis set as for the other elements are
presented.}
This basis set has proven successful in calculations of nuclear-spin
dependent $\mathcal{P}$-violating interactions and $\mathcal{P,T}$-odd
effects induced by an
eEDM in heavy polar diatomic
molecules.\cite{isaev:2012,Isaev:2013,isaev:2014,gaul:2017}
The N, F and O atoms were represented with a decontracted atomic natural
orbital (ANO) basis set of triple-$\zeta$ quality\cite{roos:2004} and for H
the s,p-subset of a decontracted correlation-consistent basis of
quadruple-$\zeta$ quality\cite{dunning:1989} was used. 
\par
The ZORA-model potential $\tilde{V}(\pos)$ was employed with
additional damping\cite{liu:2002} as proposed by van
W\"ullen\cite{wullen:1998}. In case of elements of the 8th row, the model
potential of Og, the element with highest $Z$ of all known
elements,\cite{oganessian:2006} was renormalized to the respective nuclear charge number.\par 
For the calculations of two-component wave functions and properties a
finite nucleus was used, described by a normalized spherical Gaussian
nuclear density
distribution
$\rho_A(\pos)=\rho_0\mathrm{e}^{-\frac{3}{2\zeta_A}\pos^2}$.
The root mean square radius $\zeta_A$ of nucleus $A$ was
used as suggested by Visscher and Dyall.\cite{visscher:1997}
The mass numbers $A$ were chosen as nearest integer to the standard
relative atomic mass, i.e. $^{11}$B, $^{24}$Mg, $^{27}$Al, $^{40}$Ca,
$^{45}$Sc, $^{48}$Ti, $^{65}$Zn, $^{70}$Ga, $^{88}$Sr, $^{90}$Y,
$^{91}$Zr, $^{112}$Cd, $^{115}$In, $^{137}$Ba, $^{139}$La, $^{140}$Ce,
$^{173}$Yb, $^{175}$Lu, $^{178}$Hf, $^{201}$Hg, $^{204}$Tl,
$^{226}$Ra,$^{227}$Ac, $^{232}$Th, $^{259}$No, $^{260}$Lr, $^{261}$Rf,
$^{284}$Cn; for E120 (Unbinilium, Ubn, eka-actinium) and E121
(Unbiunium, Ubu, eka-radium) the mass number was calculated by
$2.5Z$, resulting in 300 and 303, respectively.\par
The nuclear equilibrium distances were obtained at the levels of GHF-ZORA and
GKS-ZORA/B3LYP, respectively. For calculations of energy
gradients at the DFT level the nucleus was approximated as a point charge. The
distances are given in the results section.\par

\section{Results and Discussion}
\subsection{Numerical Calculation of $\mathcal{P,T}$-Violating
Properties}
In this section the study of quite a number of diatomic molecules
with $^2\Sigma_{1/2}$-ground state or for which at least a
$^2\Sigma_{1/2}$-ground can be expected, is presented, including
 group 2 mono-fluorides (Mg--E120)F, group 3 mono-oxides (Sc--E121)O , group 4 mono-nitrides
(Ti--Rf)N, group 12 mono-hydrides (Zn--Cn)H, group 13 mono-oxides
(B--Tl)O and the mono-nitrides (Ce--Th)N, mono-fluorides (Yb--No)F and
mono-oxides (Lu--Lr)O of some f-block groups, respectively. \par The
numerically calculated values of symmetry violating properties are presented for the listed molecules
together with deviations between the methods cGHF and cGKS/B3LYP in Table \ref{tab: allprops}. 
The calculated equilibrium bond length $r_\text{e}$ and numerical values of
the reduced total electronic angular momentum projection
quantum number $\Omega$ are shown as well.\par
The equilibrium bond lengths and values of $\Omega$ determined with
GHF and GKS are typically in reasonable agreement. Large deviations in
the bond length of about 0.1~$a_0$ are observed for LaO, YbF and group
13 oxides excluding BO, which indicates a more complicated
electronic structure. Nearly all values of
$\Omega$ are approximately equal to $\pm\frac{1}{2}$. Furthermore in all
cases, the reduced orbital angular momentum projection was $\Lambda\approx0$ and thus
there appears no significant contamination by $\Pi$-states. Exceptions are 
CnH and RfN as well as TiN, which show large
electron correlation effects (as gauged by the difference GHF-GKS) and seem to have a 
complicated electronic structure that requires more advanced
electronic structure methods for a reliable description. However,
even in these cases $\Lambda\approx0$ is valid and there was no significant
admixture of $\Pi$-contributions. Especially
in case of RfN the methods employed herein are not able to give reliable
results, which is indicated by enormous differences between DFT and HF
calculations, not only for properties but also for the ordering and
pairing of molecular spin-orbitals. The values given for RfN are only
included for completeness, but are not to be considered as estimates
of the expected effect sizes.\par 
Large deviations between GHF and GKS values of $W_\text{d}$
and $W_\text{s}$ can be observed for some of the group 13 oxides (esp.
AlO and GaO), which indicate that there are electron correlation effects,
which can not accurately be described by the present approaches. In
these compounds also large spin-polarization effects could be observed.
Especially for AlO more sophisticated electronic structure methods
should be applied, if more accurate results are desired. Nonetheless for the present
discussion of overall trends the description within the cGHF/cGKS scheme appears to suffice.\par
Generally the agreement between the HF and DFT descriptions is within
20~\% to 30~\%. Yet, in cases where d-orbitals play an important role,
such as group 4 nitrides or group 12 hydrides, additional electron correlation
considered via the DFT method has a pronounced impact on the value of the
$\mathcal{P,T}$-odd properties. In case of mercury mono-fluoride these
effects where already discussed in Ref. \cite{gaul:2017}.\par
The two parameters $W_\text{d}$ and $W_\text{s}$ behave analogously with
respect to inclusion of additional electron correlation effects when going
along the periodic table.\par
The largest enhancement of $\mathcal{P,T}$-odd effects can be found in
compounds of the seventh row of the periodic table, i.e. RaF, AcO, ThN,
NoF, LrO, (RfN) and CnH. But also some compounds of the sixth row show
enhancement of the similar of magnitude, namely HfN, HgH, TlO, YbF
and LuO. It shall be noted, that even the exotic molecule CnH may be
a candidate for future experiments, since ongoing research aims to achieve very
long lived isotopes for the super heavy element
Cn.\cite{oganessian:1999,oganessian:2007,utyonkov:2018}\par
The investigation of $\mathcal{P,T}$-violation in group 13 oxides
shows, that especially TlO caused problems for the methods employed
herein, as mentioned above. As comparatively large
enhancement effects were calculated for TlO, a study of this molecule
with more sophisticated electronic structure methods could
be interesting in order to obtain an accurate description of its
electronic structure. Little is known about TlO from
experimental side, however, so that significant further research would be
necessary to take advantage of such enhancement effects. 

\subsection{Estimation of $\mathcal{P,T}$-Violating Properties from
Atomic Scaling Relations\label{est}}
In order to gain deeper insight into the scaling behavior of the above
discussed properties the numerical results can be compared to
analytical and empirical atomic models. Using the relations presented in the theory
section (eqs. \prettyref{eq: wdfromwscs},\prettyref{eq: wdfromwsfs}) within the quasi-relativistic
GHF/GKS-ZORA approach the parameter $W_\text{d}$ is estimated from
$W_\text{s}$
and compared to the results of the numerical calculations.\par
Results for estimations of $W_\text{d}$ from $W_\text{s}$ for both the
analytically derived expression by Sandars and the
empirical factor found by Fermi and Segr\`e are shown in
\prettyref{tab: allest}, where again the labels FS and CS are used for
properties calculated with the corresponding factors
$\tilde{R}_\text{CS}$ and $\tilde{R}_\text{FS}$.\par
Relative deviations of the estimated $\mathcal{P,T}$-odd property
$W_\text{d}$ from the numerical calculations are typically below 10~\%
for molecules with $Z<100$.
For light molecules of the first (BO) or second row (MgF, AlO) the
deviations are much larger. In this region the atomic models do not
work well. For these cases with light elements both the analytically
derived CS-equation and the
empirical FS-relation yield much too low (BO, AlO) or too high (MgF) 
values of $W_\text{d}$. It has to be pointed out, that the case of BO is somewhat
special, since boron is even lighter than oxygen and the "heavy" atom
of this molecule is actually oxygen. By this also 
the sign of the $\mathcal{P,T}$-odd properties $W_\text{d}$ and
$W_\text{s}$ is reversed and a different
behavior than for all other group 13 compounds is expected.\par
In the region of superheavy elements ($Z>100$) the abruptly rising
analytically derived relativistic enhancement factor of the eEDM (reaching infinity at
$Z\sim118.65$) causes a large overestimation of $W_\text{d}$ resulting in
deviations of $\ge35~\%$ for NoF ($Z=102$) and LrO ($Z=103$) and 146~\% for CnH
($Z=112$) between the estimate and the numerical value. Here the
empirical factor performs much better and a much lower 
increase in the deviation from the numerical calculations can be
observed. However, even in the case of the empirically obtained relativistic
enhancement factor the $\mathcal{P,T}$-odd enhancement in superheavy
element compounds is strongly overestimated (deviations $\gg10~\%$)
with these simple atomic
models. This may be explained with the influence of the pole at $Z>137$ of the used
relativistic enhancement factors.\par
For the two studied compounds with $Z>118$ the analytically derived factor is
not applicable anymore, which results in deviations far beyond 500~\%,
whereas the estimates obtained with the empirical factor deviate still
less than 100~\%\ from numerical calculations. Nonetheless the influence of the pole at $Z=137$ of the
relativistic enhancement factors for eEDM induced permanent molecular EDMs
and scalar-pseudoscalar
nucleon-electron current interactions causes deviations $>10~\%$.

\subsection{Ratio of $\mathcal{P,T}$-violating properties} 
Various $\mathcal{P,T}$-odd parameters contribute to a permanent EDM in a
molecule. In order to set limits on more than one parameter,
experiments with different sensitivity to the $\mathcal{P,T}$-odd
parameters have to be compared. In the following we determine
the trends of the ratio of $\mathcal{P,T}$-odd enhancement
parameters in the periodic table and how the sensitivity of an
experiment to the herein discussed $\mathcal{P,T}$-odd effects
described by $d_\text{e}$ and $k_\text{s}$ is
influenced by this.\par
\begin{figure}\centering
\resizebox{.5\textwidth}{!}{\includegraphics{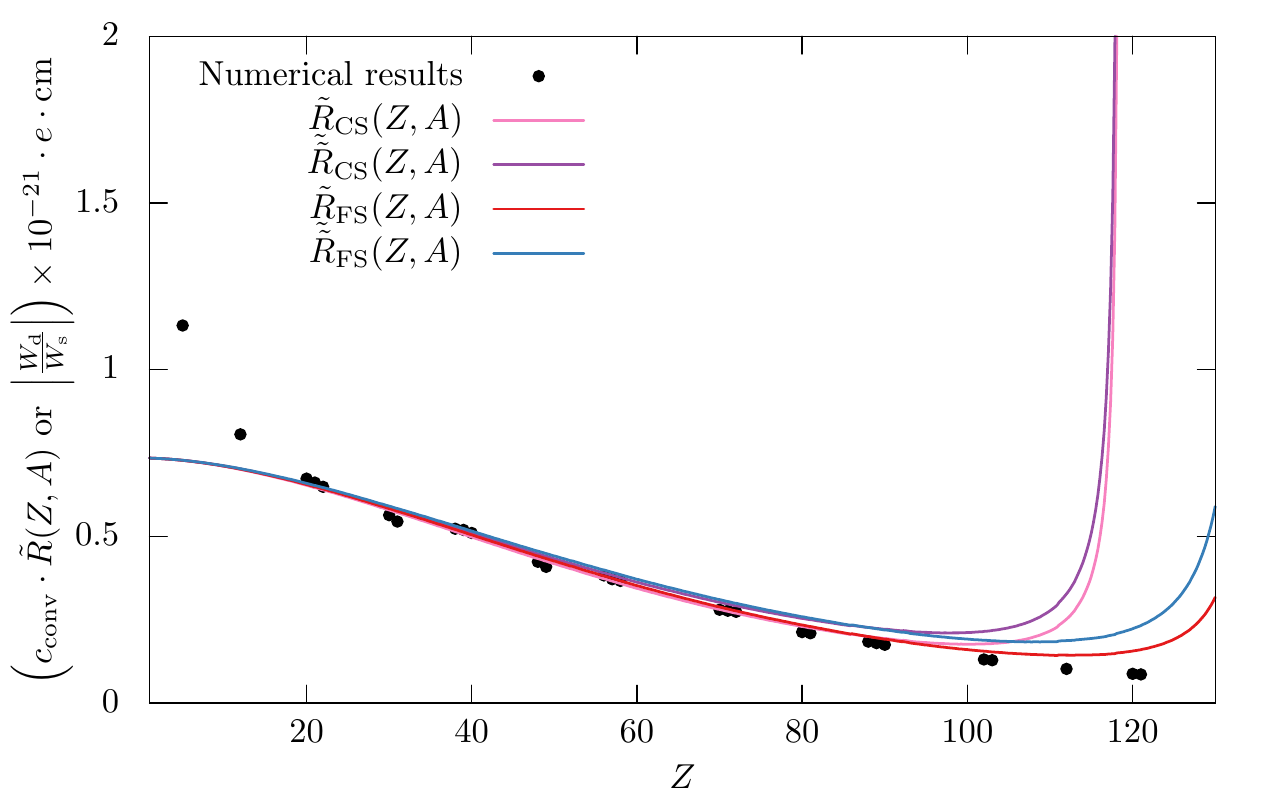}}
\caption{
Comparison of combined relativistic enhancement factors and conversion
factors for the ratio between $\mathcal{P,T}$-odd eEDM and
nucleon-electron current interactions $W_\text{d}/W_\text{s}$. The
relativistic factors $\tilde{R}$ derived from the analytically derived
factor (CS) and the empirical factor (FS) are shown, as well as there
analogs derived from an old relativistic enhancement factor for
$W_\text{s}$ $\tilde{\tilde{R}}$. Plots are shown
for the case of $j=\frac{1}{2}$ as in $^2\Sigma_{1/2}$-states.
Mass numbers $A$ were assumed as the natural mass number corresponding
to the next integer value of $Z$.
}
\label{fig: ratiofacs}
\end{figure}

The ratio $W_\text{d}/W_\text{s}$ of the various open-shell diatomic
molecules is studied, for which both the
analytically derived and the empirically derived relativistic enhancement
factors presented in section \ref{theory} are compared. 
In \prettyref{fig: ratiofacs} the ratio $W_\text{d}/W_\text{s}$
calculated with the four different relativistic enhancement factors
$\tilde{R}$ (eqs. \prettyref{eq: wdfromwscs} to \prettyref{eq: wdwsfs})
is compared to all numerical results for the value of
$W_\text{d}/W_\text{s}$. The empirically derived relativistic
enhancement factor for $W_\text{d}$ included in eqs. \prettyref{eq: wdfromwsfs} 
and \prettyref{eq: wdwsfs} is in much better agreement
with the numerical results for $Z>90$ as was also seen in the last
section in the comparison of estimates of $W_\text{d}$ with numerical
values. Furthermore values calculated with the improved relativistic enhancement factor for
$W_\text{s}$ (eq. \prettyref{eq: dzubaws}) are in better
agreement with numerical values also for $Z\ll90$.\par
However, all the ratios derived from the analytical models show a
wrong behavior in the region of $Z<30$ and $Z>90$ in comparison to the
numerical results. This causes large deviations for the estimates
discussed in the last section.\par
\begin{figure}\centering
\resizebox{.5\textwidth}{!}{\includegraphics{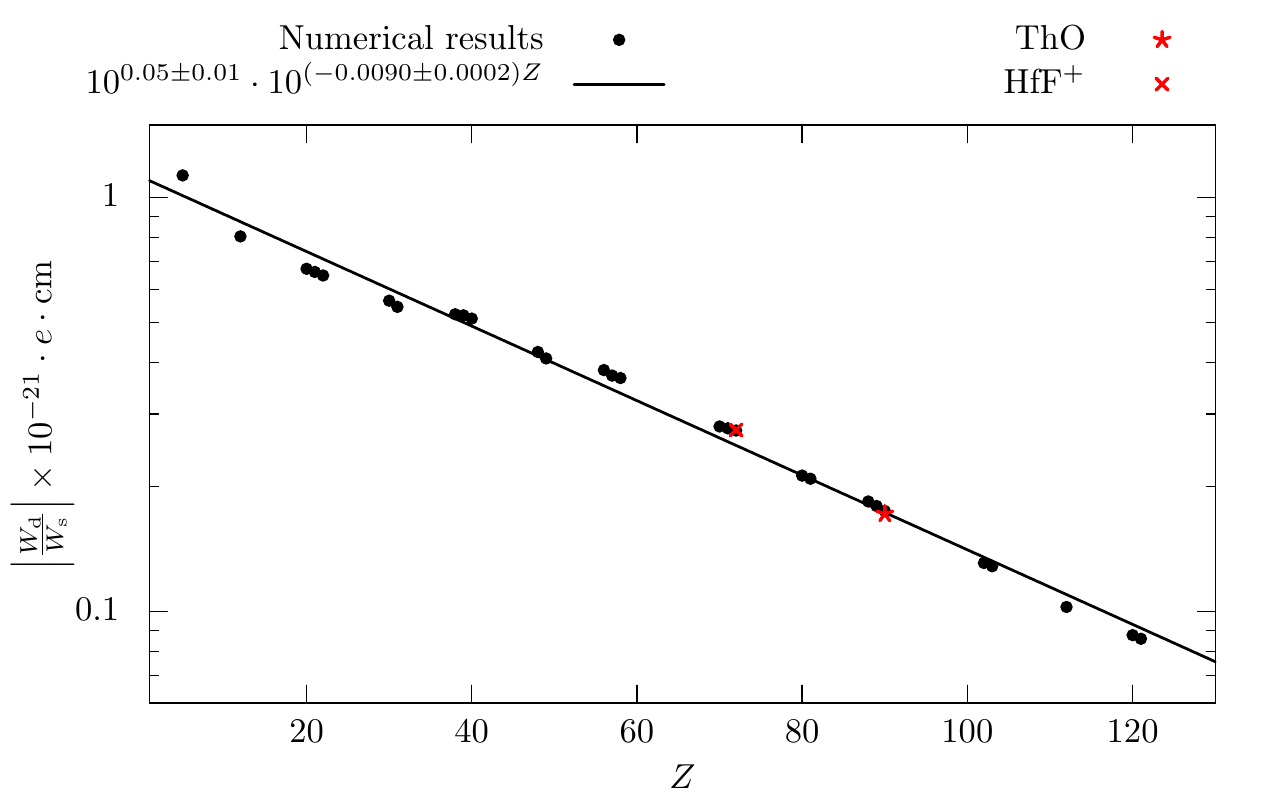}}
\caption{
Linear fit of the logarithmic $Z$-dependence of the ratio between
$\mathcal{P,T}$-odd eEDM and scalar-pseudoscalar nucleon-electron
current interactions $W_\text{d}/W_\text{s}$. The value for RfN is not
included in the fit. The values of
$W_\text{d}/W_\text{s}$ for HfF$^+$ and ThO were calculated by Fleig
in a four-component configuration interaction
framework in Ref.~\onlinecite{fleig:2017} and are shown for comparison but not included in the fit.
}
\label{fig: ratiofit}
\end{figure}

A logarithmic plot of the numerical results (see \prettyref{fig: ratiofit}) 
shows an exponential behavior of the ratio of $\mathcal{P,T}$-odd properties
$W_\text{d}/W_\text{s}$, which can be interpolated by a linear fit
model with
\begin{equation}
\log_{10}\braces{\abs{\frac{W_\text{d}}{W_\text{s}}}\times
10^{-21}~e\cdot\text{cm}}=q\cdot Z + p.
\end{equation}
In this plot in \prettyref{fig: ratiofit} also results of calculations
reported by Fleig for the two molecules, HfF$^+$ and ThO, where a
$^3\Delta$-state is of relevance for experiments, are included.\cite{fleig:2017} It can be inferred that the ratio
$W_\text{d}/W_\text{s}$ is rather insensitive to the chemical
environment of the heavy nucleus, but is essentially determined by the
exponential $Z$-dependence determined in \prettyref{fig: ratiofit}.\par
In order to disentangle the $\mathcal{P,T}$-odd parameters $k_\text{s}$ and
$d_\text{e}$ at least two experiments with molecules 1 and 2 are needed. The measurement
model than is a $2\times2$-matrix problem described by the system equations
\begin{equation}
h\begin{pmatrix}\nu_1\\\nu_2\end{pmatrix}
=\Omega\underbrace{\begin{pmatrix}
W_{\text{d},1}&W_{\text{s},1}\\
W_{\text{d},2}&W_{\text{s},2}\\
\end{pmatrix}}_{\bf C}
\begin{pmatrix}d_\text{e}\\k_\text{s}\end{pmatrix},
\end{equation}
where $\bf C$ is the matrix of sensitivity coefficients. We assume now
uncorrelated measurements with standard uncertainties $u(\nu_1)$ and
$u(\nu_2)$ and the commonly applied case of an ellipsoidal coverage
region in the parameter space of $k_\text{s}$ and
$d_\text{e}$ (for details see the Supplementary Material).The ellipse centered at
$\begin{pmatrix}d_\text{e}\\k_\text{s}\end{pmatrix}=\vec{0}$ is
described by the equation 
\begin{multline}
h^2k_p^2 =\parantheses{\frac{W_\text{d,1}^2}{u^2(\nu_1)}  
                 + \frac{W_\text{d,2}^2}{u^2(\nu_2)} } x_\text{d}^2 \\
   + 2\parantheses{\frac{W_\text{d,1}^2}{u^2(\nu_1)}  \frac{W_\text{s,1}}{W_\text{d,1}} 
                 + \frac{W_\text{d,2}^2}{u^2(\nu_2)}\frac{W_\text{s,2}}{W_\text{d,2}}}x_\text{d}x_\text{s} \\
   +  \parantheses{\frac{W_\text{d,1}^2}{u^2(\nu_1)}  \parantheses{\frac{W_\text{s,1}}{W_\text{d,1}}}^2 
                 + \frac{W_\text{d,2}^2}{u^2(\nu_2)}  \parantheses{\frac{W_\text{s,2}}{W_\text{d,2}}}^2} x_\text{s}^2
\end{multline}
where $k_p=2.45$ for an elliptical region of 95~\%
probability\cite{jcgm102:2011} and
$x_\text{d}$ and $x_\text{s}$ are the coordinates in the parameter
space in direction of $d_\text{e}$ and $k_\text{s}$, respectively. 
Thus the ellipse has an area of 
\begin{equation}
A_\text{ellipse}=
\frac{h^2k_p^2\pi\abs{u(\nu_1)u(\nu_2)}}
{\abs{W_\text{d,1}W_\text{d,2}}
\abs{\frac{W_\text{s,1}}{W_\text{d,1}} -\frac{W_\text{s,2}}{W_\text{d,2}}}}.
\end{equation}
In order two disentangle $d_\text{e}$ and $k_\text{s}$ in two
experiments and set tight limits, assuming equal uncertainties for
experiments 1 and 2 the expression 
\begin{equation}
\abs{W_{\text{d},1}W_{\text{d},2}}\,0.89\cdot\abs{1.0210^{Z_1}-1.0210^{Z_2}} \times
10^{-21}~e\cdot\text{cm}.
\end{equation}
has to become large.
The enhancement of the single experiments, which is determined by
$W_{\text{d},1}W_{\text{d},2}$ is strongly dependent on the chemical
environment, as will be discussed in the following sections. However,
assuming at this point a scaling behavior of $W_{\text{d},1}$ as in
eq. \prettyref{eq: wdscale} and eq. \prettyref{eq: wdscaleemp} for
atomic systems, the area of the coverage region is inversely proportional to
\begin{equation}
\frac{\parantheses{Z_1Z_2}^3}{\gamma^8}\,0.89\cdot\abs{1.0210^{Z_1}-1.0210^{Z_2}}
\times
10^{-21}~\frac{1}{e\cdot\text{cm}}.
\end{equation}
Thus, in order to set tight limits on both $\mathcal{P,T}$-odd
parameters, experiments with molecules that have a high nuclear charge
and at the same time differ considerably in the nuclear
charge $Z$ of the electropositive atom are required. For example when
assuming equal uncertainties $u(\nu_i)$, a comparison of experiments with YbF and
RaF or ThO would provide tighter bounds than a comparison of a BaF
experiment with a ThO experiment but also than a comparison of
experiments with  RaF and ThO. However, the
possibilities are limited for paramagnetic molecules because 
enhancement effects of the individual properties still increase
steeply with increasing $Z$, which is the dominating effect. Alternatively experiments with
diamagnetic atoms and molecules can further tighten bounds on $d_\text{e}$ and $k_\text{s}$, as they show
different dependencies on the nuclear charge (see e.g. Ref
\onlinecite{khriplovich:1997}).\par
This scheme can also be expanded for experiments that aim to set
accurate limits on more than the herein discussed parameters. However,
for this purpose first the respective enhancement factors have to be
calculated for a systematic set of molecules. Furthermore it should be
noted that the present picture is not complete because of other sources
of permanent EDMs that were not accounted for, namely
$\mathcal{P,T}$-odd tensor and pseudoscalar-scalar electron-nucleon
current interactions, as well as $\mathcal{P,T}$-odd nuclear dipole
moments, which lead to the nuclear Schiff moment and nuclear magnetic
quadrupole interactions.

\subsection{Periodic Trends of $\mathcal{P,T}$-Violating Properties}
The analytical scaling relations presented in eqs.
\prettyref{eq: dzubaws}, \prettyref{eq: wdscale} and \prettyref{eq: wdscaleemp} can also be used to determine the numerical
$Z$-scaling within a group of compounds with electropositive atoms
of the same column of the periodic table. For this purpose the property
is divided by its relativistic enhancement factor and plotted on a double logarithmic
scale, as has been done for the nuclear spin-dependent
$\mathcal{P}$-violating interaction parameter in \cite{isaev:2012,borschevsky:2013,isaev:2014}:
\begin{align}
\log_{10}&\braces{\frac{\abs{W_\text{s}}}{R(Z,A)f(Z)\frac{\gamma+1}{2}}
\times\frac{1}{h\mathrm{Hz}}
}\nonumber\\
~~~~&=b_\text{s}+\log_{10}\braces{Z^{a_\text{s}}}
\\
\log_{10}&\braces{\abs{W_\text{d}}\gamma\parantheses{4\gamma^2-1}~
\times10^{-24}\frac{e\cdot\mathrm{cm}}{h\mathrm{Hz}}
}\nonumber\\
~~~~&=b_\text{d,CS}+\log_{10}\braces{Z^{a_\text{d,CS}}}\\
\log_{10}&\braces{\abs{W_\text{d}}\gamma^4
\times10^{-24}\frac{e\cdot\mathrm{cm}}{h\mathrm{Hz}}
}\nonumber\\
~~~~&=b_\text{d,FS}+\log_{10}\braces{Z^{a_\text{d,FS}}}.
\label{eq: loglogplotwd}
\end{align} 
From eqs. (\ref{eq: wsscale}) and (\ref{eq: wdscale}) the
exponents of $Z$ can be expected to be approximately three. For both
parameters the $Z$-scaling is
studied herein not only within columns, but also for isolobal diatomics
within rows of the periodic table.\par
The resulting $Z$-exponents $a$ and factors $b$ will be
discussed in the following for both, GHF- and
GKS-results.\par

\subsubsection{$Z$-Scaling within groups of the periodic table}
\begin{figure*}\centering
\resizebox{\textwidth}{!}{\includegraphics{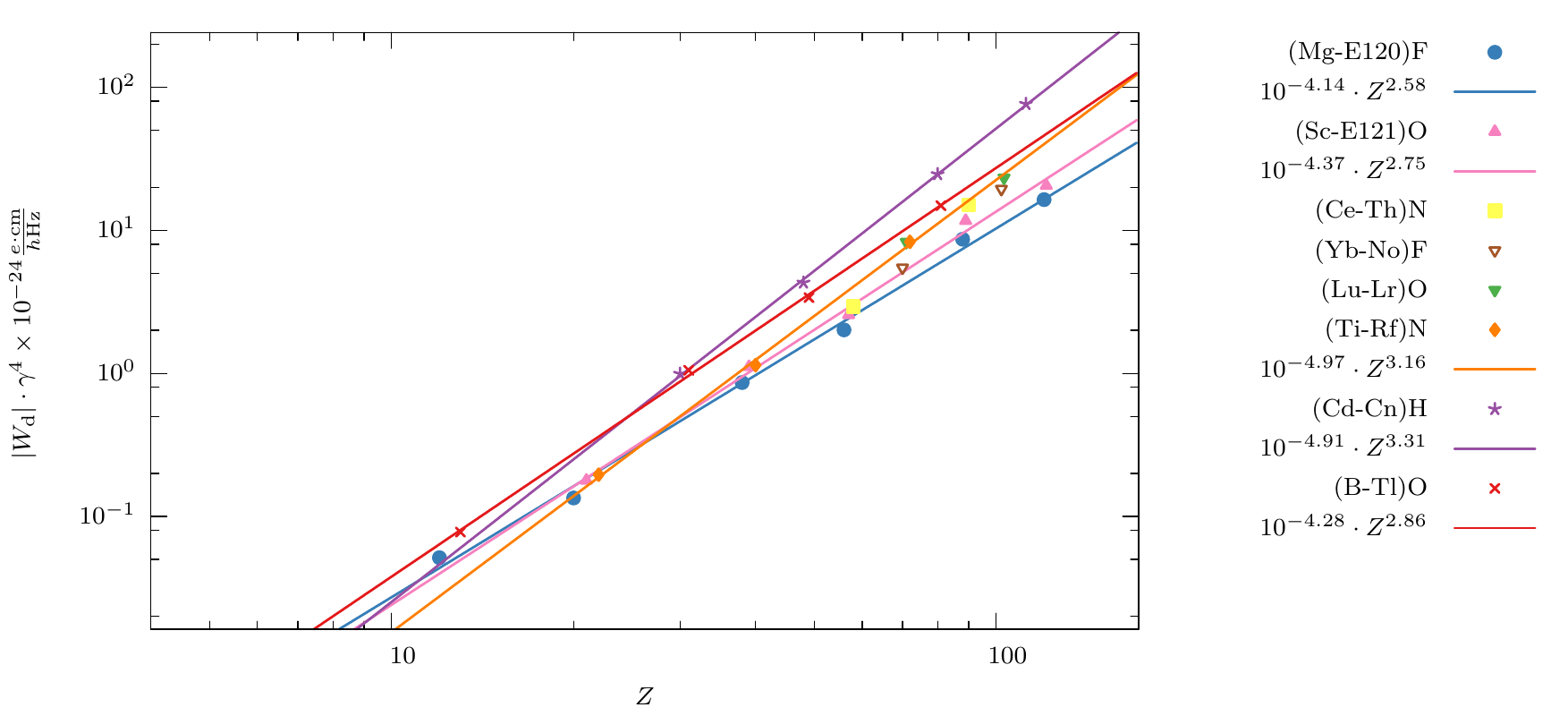}}
\resizebox{\textwidth}{!}{\includegraphics{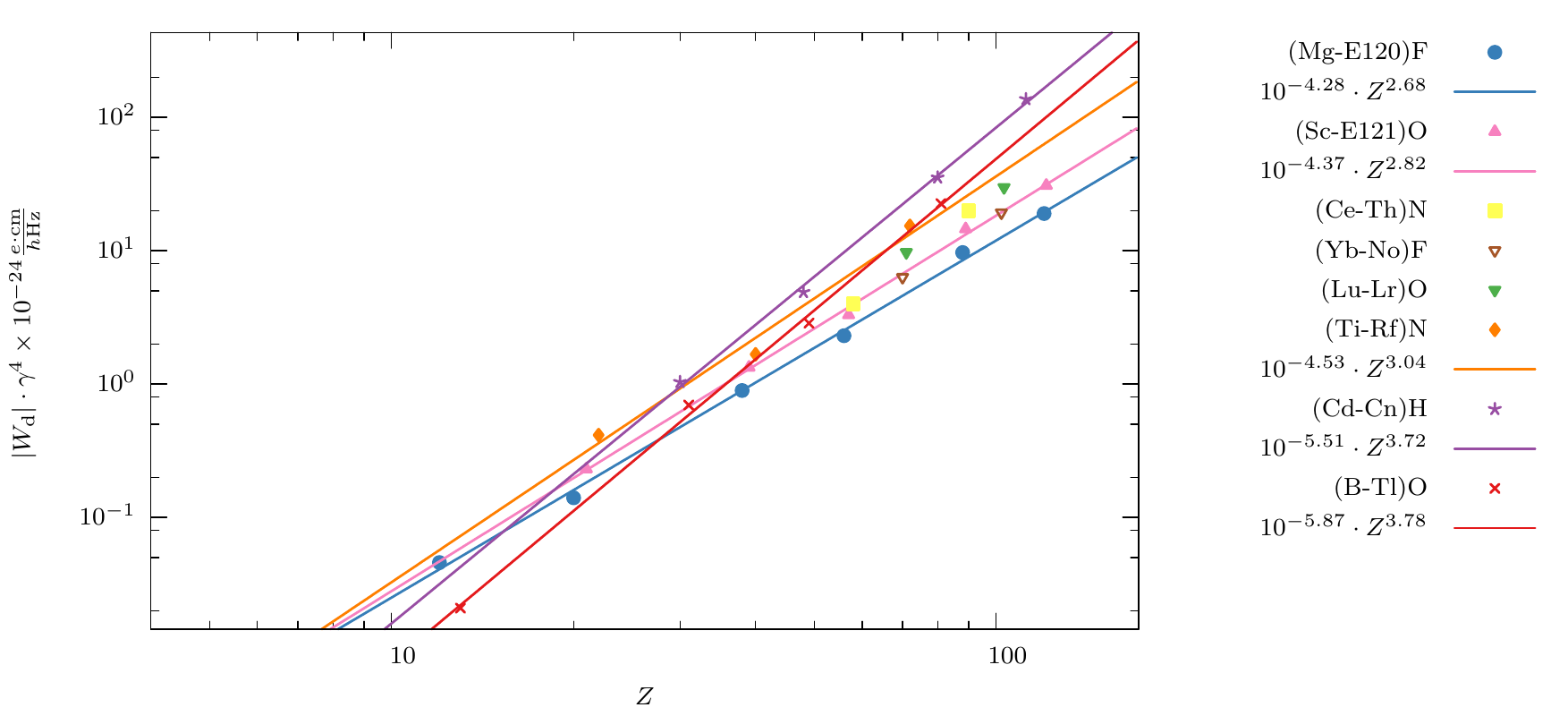}}
\caption{
Scaling of $\log_{10}\braces{|W_\text{d}|\gamma^4\times10^{-24}~\frac{e\cdot\mathrm{cm}}{h\mathrm{Hz}}}$ 
with $\log_{10}\braces{Z}$ for group 2 fluorides (Mg-E120)F, group 3
oxides (Sc-E121)O, group 4 nitrides (Ti-Hf)N, group 12 hydrides (Zn-Cn)H
and group 13 oxides (B-Tl)O at the level of GKS-ZORA/B3LYP (top) and
GHF-ZORA (bottom). The
functional expressions of the fits are assigned to the colors of the
groups. Plot of the $f$-block groups (Ce-Th)N, (Yb-No)F and (Lu-Lr)O
without fit. Boron was not included in the fit of group 13 oxides (see
text).}
\label{fig: wdtrendsgksfs}
\end{figure*}

In the following the scaling within the groups of the periodic
table is studied. The graphical representation of the $Z$-scaling
of $W_\text{s}$ and $W_\text{d}$ can be found in Figures \ref{fig:
wstrendsgks}-\ref{fig: wdtrendsgksfs}. In case of group 13
oxides, boron was not included in the linear fit, because it has a very
different character (see discussion above).\par
\begin{figure*}\centering
\resizebox{\textwidth}{!}{\includegraphics{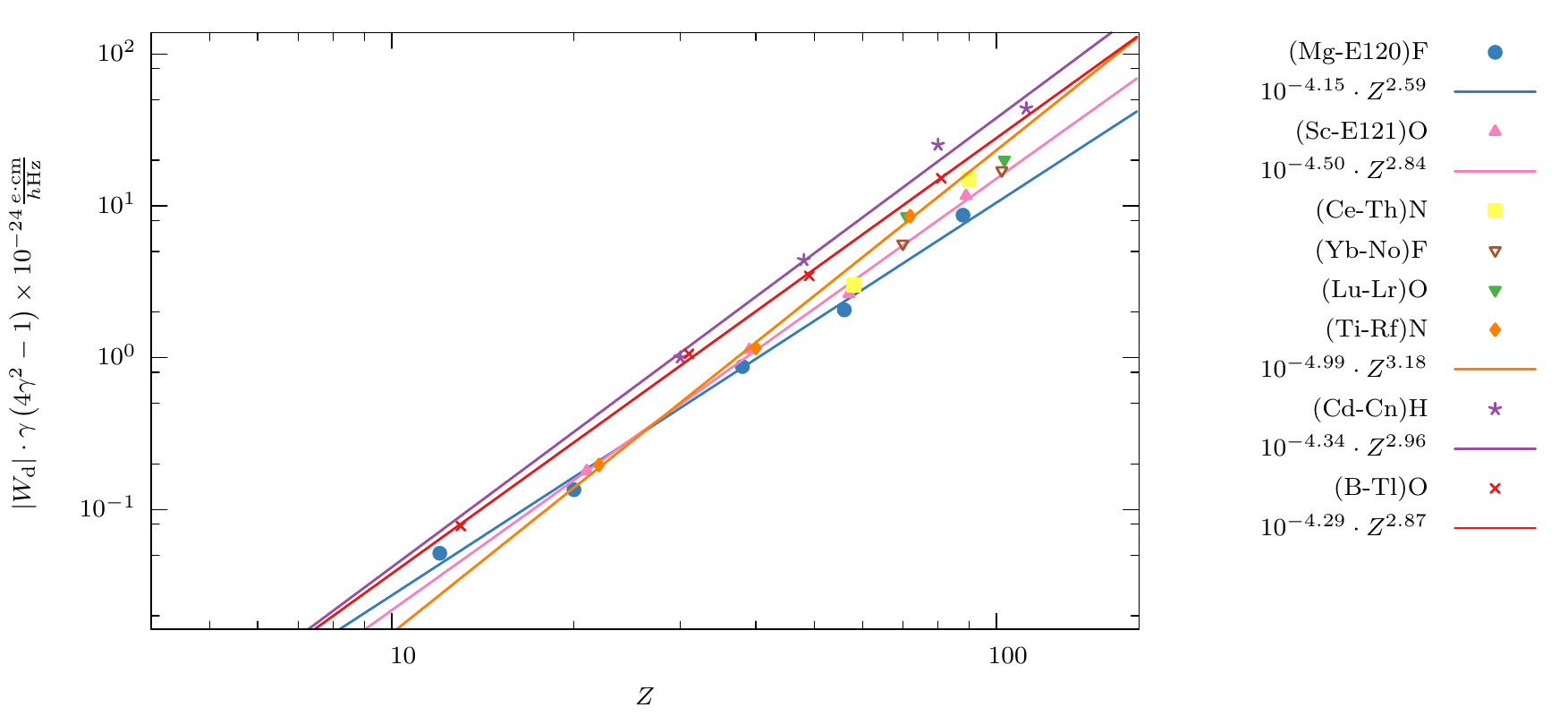}}
\resizebox{\textwidth}{!}{\includegraphics{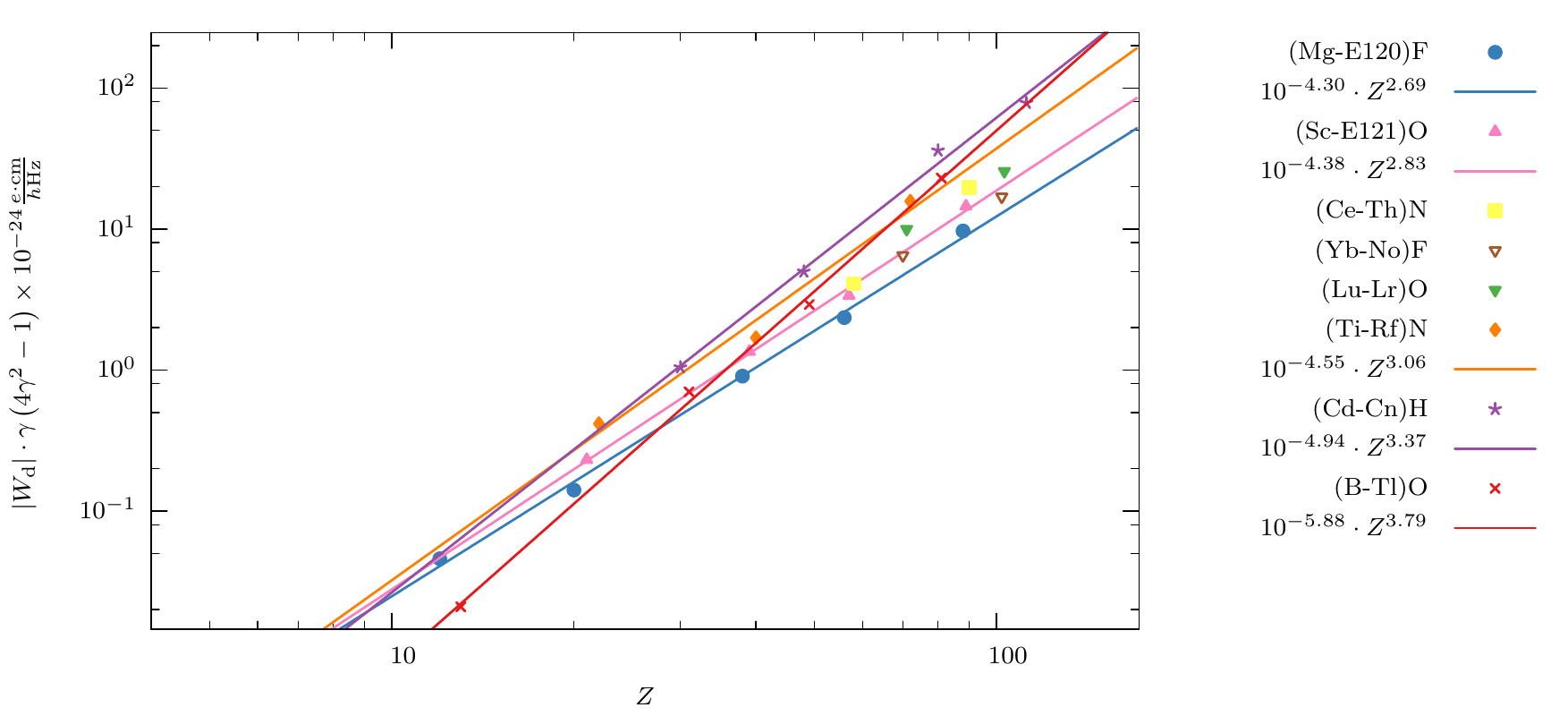}}
\caption{
Scaling of
$\log_{10}\braces{|W_\text{d}|\gamma\parantheses{4\gamma^2-1}\times10^{-24}~\frac{e\cdot\mathrm{cm}}{h\mathrm{Hz}}}$ 
with $\log_{10}\braces{Z}$ for group 2 fluorides (Mg-Ra)F, group 3
oxides (Sc-Ac)O, group 4 nitrides (Ti-Hf)N, group 12 hydrides (Zn-Cn)H
and group 13 oxides (B-Tl)O at the level of GKS-ZORA/B3LYP (top) and
GHF-ZORA (bottom). The
functional expressions of the fits are assigned to the colors of the
groups. Plot of the $f$-block groups (Ce-Th)N, (Yb-No)F and (Lu-Lr)O
without fit. Boron was not included in the fit of group 13 oxides (see
text).}
\label{fig: wdtrendsgks}
\end{figure*}

Comparing the two different relativistic enhancement factors for eEDM
interactions, which were employed in this study, we see for most
of the groups of molecules no appreciable differences between the
analytically derived and the empirical factor. Yet, in case of
group 12 hydrides it is important to use the empirical scaling factor.
Cn has a nuclear charge of $Z=112$, which is close to the singularity of
the analytically derived factor. This results in a strong
overestimation of the relativistic enhancement and thus a strong
underestimation of the plotted value, which explains the
non-linear trend for group 12 hydrides in \prettyref{fig: wdtrendsgks}. 
Furthermore with the analytically derived enhancement factor no
meaningful plot that includes the row 8 compounds E120F and E121O is
possible. Therefore in the following we will use the results
obtained with the empirical enhancement factor for our discussions.
\par
The $Z$-scaling parameters $a$ and the $Z$-independent prefactors $10^b$
are summarized in \prettyref{tab: cpvtrends}.
It should be noted, that the inclusion of the values of the row 8 compounds
into the fit causes no notable changes in the $Z$-scaling in
case of the eEDM  and $\mathcal{P,T}$-odd nucleon-electron current enhancement.\par
For nearly all parameters the agreement between GHF and GKS
calculations is excellent. The only cases, where DFT
predicts considerably different behavior, are the group 12 hydride and
group 13 oxides.
As could be seen in \cite{gaul:2017} the DFT approach performs
much better in the case of group 12 compounds than GHF due to
pronounced electron correlation effects 
and therefore can be taken as more
reliable. In the previous sections large electron correlation effects in
group 13 compounds, which lead to large differences between GHF an GKS, were already
discussed.\par
The scaling of $\mathcal{P,T}$-odd interactions  seems to follow the
same laws as that of nuclear spin-dependent
$\mathcal{P}$-violating interactions studied in
\cite{isaev:2012,borschevsky:2013}. The $Z$-scaling
increases up to group 12 hydrides, when going along the periods of
the periodic table. This maximum effect of $\mathcal{P,T}$-violation
enhancement in group 12 compounds is similar to the maximum of
relativistic and quantum electrodynamic effects in
group 11 compounds\cite{pyykko:1979,thierfelder:2010a}. At the same time the $Z$-independent
factor $10^b$ is smallest for these compounds. This damping is,
however, only dominant in the region of small $Z$, which coincides with
the findings in \cite{isaev:2012} and \cite{borschevsky:2013} for
$\mathcal{P}$-odd interactions.\par
In \cite{borschevsky:2013} the large $Z$-scaling of group 4 and group
12 compounds compared to group 2 or 3 compounds was attributed 
mainly to the filling of the d-shells, which causes an increment of
the effective nuclear charge because the shielding of the nuclear charge
by d-orbitals is less efficient than by s- or p-orbitals. Furthermore
therein it was argued that the lower
electronegativity of nitrogen compared to oxygen (group 4 shows larger
scaling than group 3, although isoelectronic) causes the large effects in
group 4 nitrides. A comparison of the molecules with f-block
elements next to group 3, that is CeN and ThN, shows a similar
behavior as for group 3 or group 2 compounds. Thus the filling of the $f$-shell has a considerable effect
on the size of $\mathcal{P,T}$-violating effects as well, which causes
group 4 nitrides to be behave differently than group 3 oxides, wheras CeN
and ThN are more similar to group 3 oxides.\par
Relating the $Z$-scaling of the fits to the expected $Z$-scaling (see
eq. \prettyref{eq: wsscale} and \prettyref{eq: wdscale}), yields a quantitative
$Z$-dependent factor for the effects of the molecular electronic structure on
$\mathcal{P,T}$-violation. Referring
to the GKS result we get an additional scaling factor of $\sim
Z^{-0.2}$ for $W_\text{s}$ and $\sim Z^{-0.4}$ for $W_\text{d}$ for group 2
fluorides, thus there is some damping of
$\mathcal{P,T}$-violating effects due to the electronic structure. This
can be observed for group 3 oxides regarding eEDM enhancement as
well ($Z^{-0.2}$ for $W_\text{d}$), but for $W_\text{s}$, in contrast, there 
is no additional $Z$-dependent damping.\par
 A similar damping can be observed for group 13 oxides on the GKS level,
whereas GHF predicts a considerable $Z$-dependent enhancement instead.
The group 4
and 12 compounds show a $Z$-dependent enhancement of
$\mathcal{P,T}$-odd effects: $\sim Z^{0.2}$ for $W_\text{s}$ and
$W_\text{d}$ in
group 4; $\sim Z^{0.5}$ for $W_\text{s}$ and $\sim Z^{0.3}$ for
$W_\text{d}$ in group 12. Thus we see a strong enhancement due to
$Z$-dependent electronic structure effects in group 12 hydrides, which
does not originate from relativistic enhancement factors obtained from atomic
considerations.\par
The $Z$-independent electronic structure factors $10^b$ show a
behavior inverse to that of $Z^a$ and are largest for group 2 fluorides and group 13
oxides in the DFT case, whereas the factors for group 12 hydrides and group 4 nitrides
are
almost an order of magnitude smaller. Yet, in GHF calculations the
$Z$-independent effects are on the same order as for group 12
hydrides. Thus, whereas the main enhancement in group 13 oxides is
$Z$-independent in the DFT description, it is $Z$-dependent in the GHF
case.\par 
Now we can return to the discussion of disentanglement of $d_\text{e}$
and $k_\text{s}$ in the two-dimensional parameter space. With the
chemical group dependent effective $Z$-dependence of the eEDM
enhancement factors for paramagnetic molecules, the area covered by two
experiments 1 and 2 in the parameter space of  $d_\text{e}$
and $k_\text{s}$ is determined by 
\begin{equation}
\frac{k_p^2 \pi \abs{u(\nu_1)u(\nu_2)}}{
10^{b_{\text{d},1}+b_{\text{d},2}}\frac{Z_1^{a_{\text{d},1}}Z_2^{a_{\text{d},2}}}{\gamma^8}~0.89\cdot\abs{1.0210^{Z_1}-1.0210^{Z_2}}
\times
10^{27}~\frac{\text{Hz}^2}{e\cdot\text{cm}}}~.
\end{equation}
Here the factor $10^{27}$ and the units result from eq.
\prettyref{eq: loglogplotwd}, wherein $W_\text{d}$ is in
units of $10^{24}\frac{h\text{Hz}}{e\cdot\text{cm}}$.\par
What remains to be analyzed in future works is the detailed influence of 
molecular orbitals on $\mathcal{P,T}$-violating effects that
causes the observed enhancement effects.\par
\subsubsection{$Z$-Scaling of isolobal molecules}
\begin{figure*}[h]\centering
\resizebox{\textwidth}{!}{\includegraphics{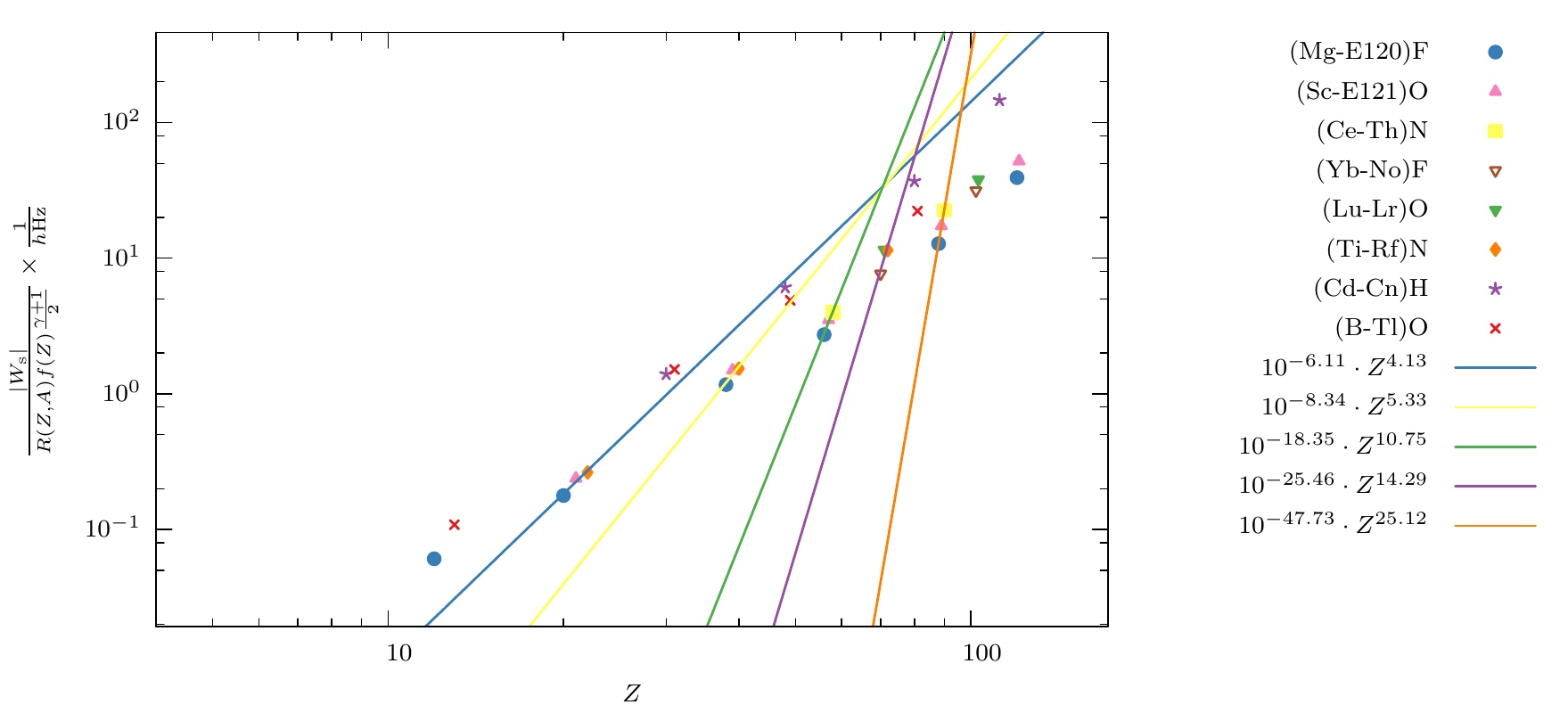}}
\resizebox{\textwidth}{!}{\includegraphics{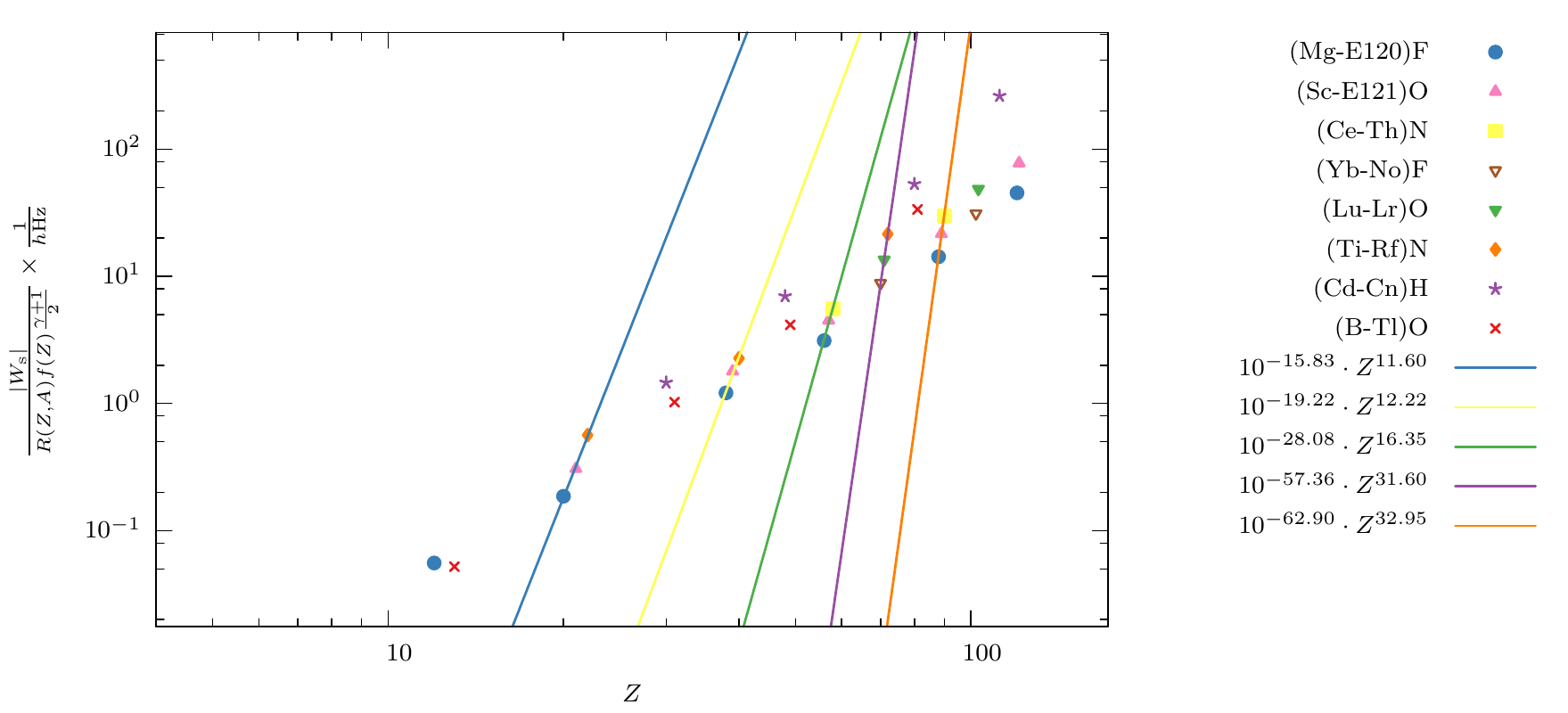}}
\caption{
Scaling of
$\log_{10}\braces{\frac{|W_\text{s}|}{R(Z,A)f(Z)\frac{\gamma+1}{2}}\times\frac{1}{h\mathrm{Hz}}}$ 
with $\log_{10}\braces{Z}$ for row 4 (Ca-Ti; blue line), row 5
(Sr-Zr; yellow line), row 6 (Ba-Ce; green line, Yb-Hf; violet line),
and row 7 (Ra-Th; organge line, No-Lr; red line) at the level of
GKS-ZORA/B3LYP (top) and
GHF-ZORA (bottom). 
}
\label{fig: wsrowtrendsgks}
\end{figure*}

Now we focus on the $Z$-scaling  for isolobal diatomic molecules
within the rows of the periodic table. When discussing eEDM
enhancement we concentrate on the results obtained with the empirical relativistic
enhancement factor in the following. For comparison, results obtained
from the analytically derived relativistic enhancement factor are
provided in the Supplemental Material. The corresponding plots can be
found in \prettyref{fig: wsrowtrendsgks} for $W_\text{s}$ and
\prettyref{fig:
wdrowtrendsgks} for $W_\text{d}$ and the resulting scaling and damping
parameters are listed in \prettyref{tab: cpvrowtrends}.\par
Trends, similar to those reported in \cite{isaev:2014} for
the $\mathcal{P}$-odd nucelar spin-dependent interaction can also be
observed for the $\mathcal{P,T}$-odd properties. However, we can see a large discrepancy
between results obtained from GHF and GKS calculations. Big
deviations between the GHF and GKS results in the fourth and fifth row
probably stem from electron correlation effects, which lead to a
considerable reduction of the $Z$-scaling, in group 6 compounds. Fits of the
DFT results have large errors that lead to qualitative differences.
Especially for row 6 compounds with a filled f-shell (violet line in
\prettyref{fig: wsrowtrendsgks} and \prettyref{fig:
wdrowtrendsgks}) a large fit error can be observed, since HfN does not
fold into a linear fit model. The results of GHF fit much better
into the trend and show that the scaling behavior of post-f-block
compounds of row 6 is approximately similar to that of row 7 compounds
without a filled f-shell. Comparing compounds with a filled d-shell
(group 12 and 13), we see that the slope becomes negative.
This again indicates a maximum of enhancement of $\mathcal{P,T}$-odd
effects in group 12 as discussed before.\par
 The investigations show that the chemical
environment of the heavy atom can have a much more important effect on
the $Z$-dependent enhancement than the physical nature of the atom.
This can result in effects scaling as $\sim Z^{30}$ for row 7 compounds. Thus a more
complex chemical environment may allow for better tuning of the size
of $\mathcal{P,T}$-odd enhancement effects. Hence we may speculate
that polyatomic molecules might be capable to give larger enhancement effects due to the
electronic structure surrounding the heavy atom.
\par

\section{Conclusion}
In this paper we calculated $\mathcal{P,T}$-odd properties due to eEDM
and nucleon-electron current interactions in polar open-shell diatomic
molecules. We determined periodic trends of $\mathcal{P,T}$-violation
by comparison to atomic scaling relations and showed that the trends are very
similar to those of nuclear spin-dependent $\mathcal{P}$-violating
interactions. Furthermore this comparison revealed problems of
frequently used scaling relation for eEDM enhancement in the regime
of heavy elements with $Z>100$. We showed that an alternative
relativistic enhancement factor found empirically by Fermi and Segr\`e
can resolve partially the problems for $Z<137$. Group 12 hydrides and group 4 nitrides were
identified to show a very steep $Z$-scaling and therefore interesting
$Z$-dependent electronic structure effects, enhancing
$\mathcal{P,T}$-violation in these compounds, were
identified. Furthermore, a study of the ratio between $\mathcal{P,T}$-odd
properties $W_\text{d}/W_\text{s}$, showed that electronic structure
effects and the chemical environment have a very low influence on the
ratio, and the ratio is mainly determined by an exponential dependence on the
nuclear charge $Z$. Thus for experiments which aim to differentiate
between $d_\text{e}$ and
$k_\text{s}$, the use of molecules with a relatively large difference in
nuclear charge $Z$ would be favorable.
The analysis of the scaling of isolobal systems and the study of the
ratio $W_\text{d}/W_\text{s}$ showed the
limitations of polar diatomic molecules and points to possible advantages in the
use of more complex systems, such as polyatomic molecules. The latter
will be focus of future research in our lab. 

\section{Supplemental Material}
See the Supplemental Material for details on the used basis sets,
further plots of trends derived with the analytical relativistic
enhancement factor by Sandars and a comparison of results received
from alternative forms of the eEDM interaction Hamiltonian.
\begin{acknowledgments}
Financial support by the State Initiative for the Development of Scientific
and Economic Excellence (LOEWE) in the LOEWE-Focus ELCH and computer time
provided by the center for scientific computing (CSC) Frankfurt are
gratefully acknowledged. T.I. is grateful to RFBR grant N 16-02-01064 for
partial support. S.M. gratefully acknowledges support from Fonds der
Chemischen Industrie. We thank Yuri Oganessian for inspiring discussions on
super heavy elements.
\end{acknowledgments}

%
\clearpage

\begin{table*}\centering
\begin{threeparttable}
\caption{Diatomic constants and $\mathcal{P,T}$-violating properties
of diatomic molecules calculated \textit{ab initio} within a
quasi-relativsitc two-component ZORA approach at the cGHF and
cGKS/B3LYP level. Dev. refers to the relative deviation between cGHF
and cGKS results.}
\label{tab: allprops}
\begin{tabular}{
l
S[table-number-alignment=center,table-format=3.0]
S[table-number-alignment=center,table-format=2.2,round-precision=2,round-mode=places]
S[table-number-alignment=center,table-format=2.2,round-precision=2,round-mode=places]
c
S[table-number-alignment=center,table-format=3.3,round-precision=3,round-mode=places]
S[table-number-alignment=center,table-format=3.3,round-precision=3,round-mode=places]
c
S[table-number-alignment=center,table-format=3.2,table-figures-exponent=2,round-precision=2,round-mode=places]
S[table-number-alignment=center,table-format=3.2,table-figures-exponent=2,round-precision=2,round-mode=places]
S[table-number-alignment=center,table-format=2.0,round-precision=0,round-mode=places]<{{\si{\percent}}}
c
S[table-number-alignment=center,table-format=3.2,table-figures-exponent=2,round-precision=2,round-mode=places]
S[table-number-alignment=center,table-format=3.2,table-figures-exponent=2,round-precision=2,round-mode=places]
S[table-number-alignment=center,table-format=3.0,round-precision=0,round-mode=places]<{{\si{\percent}}}
}
\toprule
Molecule& {$Z$}& 
\multicolumn{2}{c}{$r_\text{e}/a_0$}&
&
\multicolumn{2}{c}{$\Omega$\tnote{**}}&
&
\multicolumn{3}{c}{$W_\text{s}\frac{1}{h\cdot\text{Hz}}$}&
&
\multicolumn{3}{c}{$W_\text{d}\frac{e\cdot\text{cm}}{10^{24}\cdot
h\cdot\text{Hz}}$}\\
\cline{3-4}
\cline{6-7}
\cline{9-11}
\cline{13-15}
&&
cGHF&cGKS&&
cGHF&cGKS&&
cGHF&cGKS&\multicolumn{1}{c}{Dev.}&&
cGHF&cGKS&\multicolumn{1}{c}{Dev.}\\
\midrule
\multicolumn{15}{c}{group 2 fluorides}\\
MgF &  12 &  3.28 & 3.33 &&  0.5000 &  0.5000      &        & -5.9262e+01 & -6.4806e+01 &  9.35 & &-4.6562e-02 & -5.2244e-02              & 12.20\\
CaF &  20 &  3.74 & 3.68 &&  0.5000 &  0.5000      &        & -2.1858e+02 & -2.0864e+02 &  4.55 & &-1.4656e-01 & -1.4046e-01              &  4.17\\
SrF &  38 &  3.98 & 3.94 && -0.5000 &  0.5000      &        & -2.0096e+03 & -1.9383e+03 &  3.55 & &-1.0494e+00 & -1.0132e+00              &  3.45\\
BaF &  56 &  4.16 & 4.11 &&  0.4999 &  0.4999      &        & -8.6713e+03 & -7.5857e+03 & 12.52 & &-3.3201e+00 & -2.9051e+00              & 12.50\\
RaF &  88 &  4.30 & 4.26 && -0.4996 & -0.4995      &        & -1.5218e+05 & -1.3634e+05 & 10.41 & &-2.8056e+01 & -2.5126e+01              & 10.44\\
E120F &  120 &  4.37 & 4.35 && 0.5000 & 0.4988      &        &
-3.98133458018088e6 & -3.44698437817762e6 & 13.42 & &-3.494151249566e2 &
-3.017813922259e2              & 13.63\\
\multicolumn{15}{c}{group 3 oxides}\\                                                         
ScO &  21 &  3.15 & 3.14 &&  0.5001 &  0.5000      &        & -3.6488e+02 & -2.8309e+02 & 22.42 & &-2.4187e-01 & -1.8725e-01              & 22.58\\
 YO &  39 &  3.37 & 3.39 &&  0.5002 &  0.5001      &        & -3.0419e+03 & -2.5352e+03 & 16.66 & &-1.5812e+00 & -1.3168e+00              & 16.72\\
LaO &  57 &  3.60 & 3.46 &&  0.4998 &  0.4999      &        & -1.2957e+04 & -1.0112e+04 & 21.96 & &-4.8235e+00 & -3.7569e+00              & 22.11\\
AcO &  89 &  3.64 & 3.67 &&  0.4981 & -0.4989      &        & -2.4276e+05 & -1.9443e+05 & 19.91 & &-4.3597e+01 & -3.4940e+01              & 19.86\\
E121O &  121 &  3.82 & 3.87 && 0.5000 & -0.5000      &        &
-7.40754404594871e6 & -4.93868060321056e6 & 33.33 & &-6.358740880388e2 &
-4.237932507310e2             & 33.35\\
\multicolumn{15}{c}{group 4 nitrides}\\                                                           
TiN &  22 &  2.94 & 2.94 &&  0.3579 &  0.3579      &        & -6.8055e+02 & -3.1813e+02 & 53.25 & &-4.3682e-01 & -2.0631e-01              & 52.77\\
ZrN &  40 &  3.11 & 3.19 &&  0.4923 &  0.4920      &        & -3.9512e+03 & -2.6751e+03 & 32.30 & &-1.9983e+00 & -1.3649e+00              & 31.70\\
HfN &  72 &  3.30 & 3.26 &&  0.5008 &  0.4988      &        & -1.0917e+05 & -5.8062e+04 & 46.82 & &-2.9293e+01 & -1.5886e+01              & 45.77\\
RfN\tnote{*} & 104 &  {(}3.55{)} &{(}3.48{)} &&{(}-0.4106{)} 
&{(}-0.4764{)}&&{(}2.4847e+06{)} & {(}1.6844e+05{)}  & 93.22 &&{(}3.0602e+02{)} & {(}1.7891e+01{)}            & 94.15\\
\multicolumn{15}{c}{f-block nitrides}\\                                                       
CeN &  58 &  3.29 & 3.26 &&  0.4992 &  0.4980      &        & -1.6482e+04 & -1.1846e+04 & 28.13 && -5.9513e+00 & -4.3401e+00              & 27.07\\
ThN &  90 &  3.41 & 3.44 &&  0.4948 &  0.4968      &        & -3.5402e+05 & -2.6602e+05 & 24.86 && -6.1628e+01 & -4.6494e+01              & 24.56\\
\multicolumn{15}{c}{f-block fluorides}\\                                                      
YbF &  70 &  3.90 & 3.76 &&  0.4997 &  0.4734      &        & -4.1259e+04 & -3.5770e+04 & 13.30 && -1.1570e+01 & -1.0012e+01              & 13.47\\
NoF & 102 &  3.96 & 3.92 &&  0.4981 & -0.4937      &        & -7.3985e+05 & -7.4672e+05 &  0.93 && -9.6906e+01 & -9.7692e+01              &  0.81\\
\multicolumn{15}{c}{f-block oxides}\\                                                         
LuO &  71 &  3.41 & 3.39 &&  0.4997 &  0.4996      &        & -6.5692e+04 & -5.5908e+04 & 14.89 && -1.8154e+01 & -1.5482e+01              & 14.72\\
LrO & 103 &  3.51 & 3.53 && -0.4949 & -0.4893      &        & -1.2287e+06 & -9.5842e+05 & 22.00 && -1.5750e+02 & -1.2319e+02              & 21.78\\
\multicolumn{15}{c}{group 12 hydrides}\\                                                      
ZnH &  30 &  3.05 & 3.04 && -0.4999 & -0.4999      &        & -2.0282e+03 & -1.9429e+03 &  4.21 & &-1.1421e+00 & -1.0951e+00              &  4.12\\
CdH &  48 &  3.36 & 3.38 &&  0.4994 &  0.4995      &        & -1.5098e+04 & -1.3219e+04 & 12.45 & &-6.3611e+00 & -5.5994e+00              & 11.97\\
HgH &  80 &  3.30 & 3.33 &&  0.4907 &  0.4924      &        & -3.8455e+05 & -2.6720e+05 & 30.52 & &-8.1335e+01 & -5.6866e+01              & 30.08\\
CnH & 112 &  3.04 & 3.13 &&  0.3498 & -0.3879      &        & -1.2169e+07 & -6.7749e+06 & 44.33 & &-1.2422e+03 & -6.9398e+02              & 44.13\\
\multicolumn{15}{c}{group 13 oxides}\\                                                        
 BO &   5 &  2.23 & 2.27 && -0.5000 & -0.5000     &         &8.885122391  &  9.3123e+00 &  4.81 &&  9.4246e-03 & 1.0546e-02               & 11.90\\
AlO &  13 &  3.17 & 3.07 &&  0.4997 &  0.5000     &         & -5.5985e+01 & -1.1679e+02 &108.61 && -2.1279e-02 & -7.9068e-02              &271.58\\
GaO &  31 &  3.37 & 3.24 &&  0.4999 &  0.4999     &         & -1.4555e+03 & -2.1463e+03 & 47.46 && -7.7262e-01 & -1.1688e+00              & 39.62\\
InO &  49 &  3.79 & 3.67 && -0.4988 & -0.4990     &         & -9.2773e+03 & -1.0902e+04 & 17.51 && -3.7595e+00 & -4.4550e+00              & 18.50\\
TlO &  81 &  4.09 & 3.86 &&  0.4828 &  0.4859     &         & -2.5439e+05 & -1.6821e+05 & 33.88 && -5.3361e+01 & -3.5187e+01              & 34.10\\
\bottomrule
\end{tabular}
\begin{tablenotes}
\item[*] No reliable results could be obtained for RfN.
\item[**] The absolute sign of $\Omega$ is arbitrary. However, relative
to the sign of the effective electric field $W_\text{d}\Omega$ it is
always such that $\text{sgn}\parantheses{W_\text{d}}=-1$. Exceptions from
this (RfN and BO) are discussed in the text.
\end{tablenotes}
\end{threeparttable}
\end{table*}
%
\begin{table*}\centering
\begin{threeparttable}
\caption{eEDM enhancement parmeter $W_\text{d}$ of diatomic molecules
estimated from numerically calculated $\mathcal{P,T}$-odd interaction
parameter $W_\text{s}$ via an analytical and an empirical relation
from atomic considerations and comparison to numerical results.
$\Delta_\text{CS/FS}=\left|\frac{W_\text{d}-W_\text{d,CS/FS}}{W_\text{d}}\right|$
refers to the relative deviation of estimates with respect to
numerical calculations.}
\label{tab: allest}
\begin{tabular}{
l
S[table-number-alignment=center,table-format=3.0]
S[table-number-alignment=center,table-format=3.1,table-figures-exponent=2,round-precision=1,round-mode=places]
S[table-number-alignment=center,table-format=3.0,round-precision=0,round-mode=places,scientific-notation= fixed, fixed-exponent = 0]<{{\si{\percent}}}
S[table-number-alignment=center,table-format=3.1,table-figures-exponent=2,round-precision=1,round-mode=places]
S[table-number-alignment=center,table-format=3.0,round-precision=0,round-mode=places,scientific-notation= fixed, fixed-exponent = 0]<{{\si{\percent}}}
c
S[table-number-alignment=center,table-format=3.1,table-figures-exponent=2,round-precision=1,round-mode=places]
S[table-number-alignment=center,table-format=3.0,round-precision=0,round-mode=places,scientific-notation= fixed, fixed-exponent = 0]<{{\si{\percent}}}
S[table-number-alignment=center,table-format=3.1,table-figures-exponent=2,round-precision=1,round-mode=places]
S[table-number-alignment=center,table-format=3.0,round-precision=0,round-mode=places,scientific-notation= fixed, fixed-exponent = 0]<{{\si{\percent}}}
}
\toprule
&&
\multicolumn{4}{c}{cGHF}&&
\multicolumn{4}{c}{cGKS}
\\
\cline{3-6}
\cline{8-11}
Molecule& {$Z$} &
{$W_\text{d,CS}\frac{e\cdot\text{cm}}{10^{24}\cdot h\cdot\text{Hz}}$}&
\multicolumn{1}{c}{$\Delta_{\text{CS}}$}&
{$W_\text{d,FS}\frac{e\cdot\text{cm}}{10^{24}\cdot h\cdot\text{Hz}}$}&
\multicolumn{1}{c}{$\Delta_{\text{FS}}$}&
&
{$W_\text{d,CS}\frac{e\cdot\text{cm}}{10^{24}\cdot h\cdot\text{Hz}}$}&
\multicolumn{1}{c}{$\Delta_{\text{CS}}$}&
{$W_\text{d,FS}\frac{e\cdot\text{cm}}{10^{24}\cdot h\cdot\text{Hz}}$}&
\multicolumn{1}{c}{$\Delta_{\text{FS}}$}\\
\midrule
\multicolumn{11}{c}{group 2 fluorides}\\
MgF &  12                & -4.1534e-02 & 1.0799e+01 & -4.1587e-02 &
1.0685e+01  &            & -4.5419e-02 & 1.3064e+01 & -4.5477e-02 &
1.2954e+01\\
CaF &  20                & -1.4277e-01 & 2.5866e+00 & -1.4327e-01 &
2.2466e+00  &            & -1.3628e-01 & 2.9748e+00 & -1.3675e-01 &
2.6361e+00\\
SrF &  38                & -1.0343e+00 & 1.4373e+00 & -1.0466e+00 &
2.6271e-01 &             & -9.9759e-01 & 1.5404e+00 & -1.0095e+00 &
3.6708e-01\\
BaF &  56                & -3.2369e+00 & 2.5073e+00 & -3.3095e+00 &
3.1914e-01 &             & -2.8317e+00 & 2.5282e+00 & -2.8952e+00 &
3.4051e-01\\
RaF &  88                & -3.0409e+01 & 8.3852e+00 & -3.0389e+01 &
8.3159e+00 &             & -2.7244e+01 & 8.4303e+00 & -2.7226e+01 &
8.3610e+00\\
E120F & 120                & 3.0786e+03 & 9.8107e+02 & -6.1314e+02 &
7.5477e+01   &           & 2.6654e+03 & 9.8322e+02 & -5.3085e+02 &
7.5906e+01\\

\multicolumn{11}{c}{group 3 oxides}\\                                                         
ScO &  21                & -2.3594e-01 & 2.4491e+00 & -2.3685e-01 &
2.0744e+00 &             & -1.8306e-01 & 2.2425e+00 & -1.8376e-01 &
1.8670e+00\\
 YO &  39                & -1.5401e+00 & 2.6020e+00 & -1.5593e+00 &
1.3853e+00 &             & -1.2836e+00 & 2.5249e+00 & -1.2996e+00 &
1.3074e+00\\
LaO &  57                & -4.7405e+00 & 1.7207e+00 & -4.8493e+00 &
5.3540e-01 &             & -3.6995e+00 & 1.5290e+00 & -3.7844e+00 &
7.3149e-01\\
AcO &  89                & -4.7705e+01 & 9.4237e+00 & -4.7482e+01 &
8.9107e+00 &             & -3.8207e+01 & 9.3507e+00 & -3.8028e+01 &
8.8381e+00\\
E121O & 121                & 3.0661e+03 & 5.8218e+02 & -1.1720e+03 &
8.4313e+01   &           & 2.0442e+03 & 5.8235e+02 & -7.8138e+02 &
8.4379e+01\\

\multicolumn{11}{c}{group 4 nitrides}\\                                                           
TiN &  22                & -4.3536e-01 & 3.3326e-01 & -4.3719e-01 &
8.6041e-02 &             & -2.0351e-01 & 1.3574e+00 & -2.0437e-01 &
9.4240e-01\\
ZrN &  40                & -1.9680e+00 & 1.5153e+00 & -1.9937e+00 &
2.2781e-01  &            & -1.3324e+00 & 2.3786e+00 & -1.3498e+00 &
1.1024e+00\\
HfN &  72                & -2.9443e+01 & 5.1151e-01 & -3.0205e+01 &
3.1155e+00 &             & -1.5658e+01 & 1.4300e+00 & -1.6064e+01 &
1.1237e+00\\
RfN\tnote{*}  & 104                & {(}4.4862e+02 {)}& 4.6598e+01
& {(}3.7701e+02{)} & 2.3197e+01  &            & {(}3.0412e+01{)} &
6.9982e+01 & {(}2.5557e+01{)} & 4.2848e+01\\
\multicolumn{11}{c}{f-block nitrides}\\                                                       
CeN &  58                & -5.9082e+00 & 7.2486e-01 & -6.0467e+00 &
1.6027e+00 &             & -4.2464e+00 & 2.1571e+00 & -4.3460e+00 &
1.3685e-01\\
ThN &  90                & -6.8658e+01 & 1.1408e+01 & -6.8027e+01 &
1.0383e+01 &             & -5.1591e+01 & 1.0963e+01 & -5.1117e+01 &
9.9427e+00\\
\multicolumn{11}{c}{f-block fluorides}\\                                                      
YbF &  70                & -1.1592e+01 & 1.9087e-01 & -1.1897e+01 &
2.8276e+00  &            & -1.0050e+01 & 3.8510e-01 & -1.0315e+01 &
3.0269e+00\\
NoF & 102                & -1.3130e+02 & 3.5494e+01 & -1.1520e+02 &
1.8875e+01 &             & -1.3252e+02 & 3.5654e+01 & -1.1627e+02 &
1.9016e+01\\
\multicolumn{11}{c}{f-block oxides}\\                                                         
LuO &  71                & -1.8075e+01 & 4.3436e-01 & -1.8547e+01 &
2.1701e+00 &             & -1.5383e+01 & 6.4227e-01 & -1.5785e+01 &
1.9567e+00\\
LrO & 103                & -2.1954e+02 & 3.9392e+01 & -1.8877e+02 &
1.9857e+01 &             & -1.7125e+02 & 3.9010e+01 & -1.4725e+02 &
1.9528e+01\\
\multicolumn{11}{c}{group 12 hydrides}\\                                                      
ZnH &  30                & -1.1746e+00 & 2.8459e+00 & -1.1836e+00 &
3.6341e+00 &             & -1.1252e+00 & 2.7493e+00 & -1.1338e+00 &
3.5368e+00\\
CdH &  48                & -6.5467e+00 & 2.9178e+00 & -6.6637e+00 &
4.7567e+00 &             & -5.7319e+00 & 2.3673e+00 & -5.8343e+00 &
4.1964e+00\\
HgH &  80                & -8.8394e+01 & 8.6790e+00 & -9.0109e+01 &
1.0788e+01 &             & -6.1419e+01 & 8.0064e+00 & -6.2611e+01 &
1.0103e+01\\
CnH & 112                & -3.0614e+03 & 1.4645e+02 & -1.7499e+03 &
4.0872e+01 &             & -1.7044e+03 & 1.4560e+02 & -9.7425e+02 &
4.0386e+01\\
\multicolumn{11}{c}{group 13 oxides}\\                                                        
 BO &   5                & 6.4678e-03 & 3.1374e+01 & 6.4692e-03 &
3.1359e+01  &            & 6.7787e-03 & 3.5720e+01 & 6.7802e-03 &
3.5706e+01\\
AlO &  13                & -3.8953e-02 & 8.3062e+01 & -3.9011e-02 &
8.3335e+01 &             & -8.1263e-02 & 2.7761e+00 & -8.1384e-02 &
2.9292e+00\\
GaO &  31                & -8.3176e-01 & 7.6534e+00 & -8.3854e-01 &
8.5315e+00 &             & -1.2265e+00 & 4.9359e+00 & -1.2365e+00 &
5.7918e+00\\
InO &  49                & -3.9495e+00 & 5.0528e+00 & -4.0224e+00 &
6.9929e+00 &             & -4.6412e+00 & 4.1800e+00 & -4.7269e+00 &
6.1040e+00\\
TlO &  81                & -5.7406e+01 & 7.5793e+00 & -5.8430e+01 &
9.4987e+00 &             & -3.7957e+01 & 7.8739e+00 & -3.8634e+01 &
9.7985e+00\\

\bottomrule
\end{tabular}
\begin{tablenotes}
\item[*] No reliable results could be obtained for RfN.
\end{tablenotes}
\end{threeparttable}
\end{table*}
%


\begin{figure*}\centering
\resizebox{\textwidth}{!}{\includegraphics{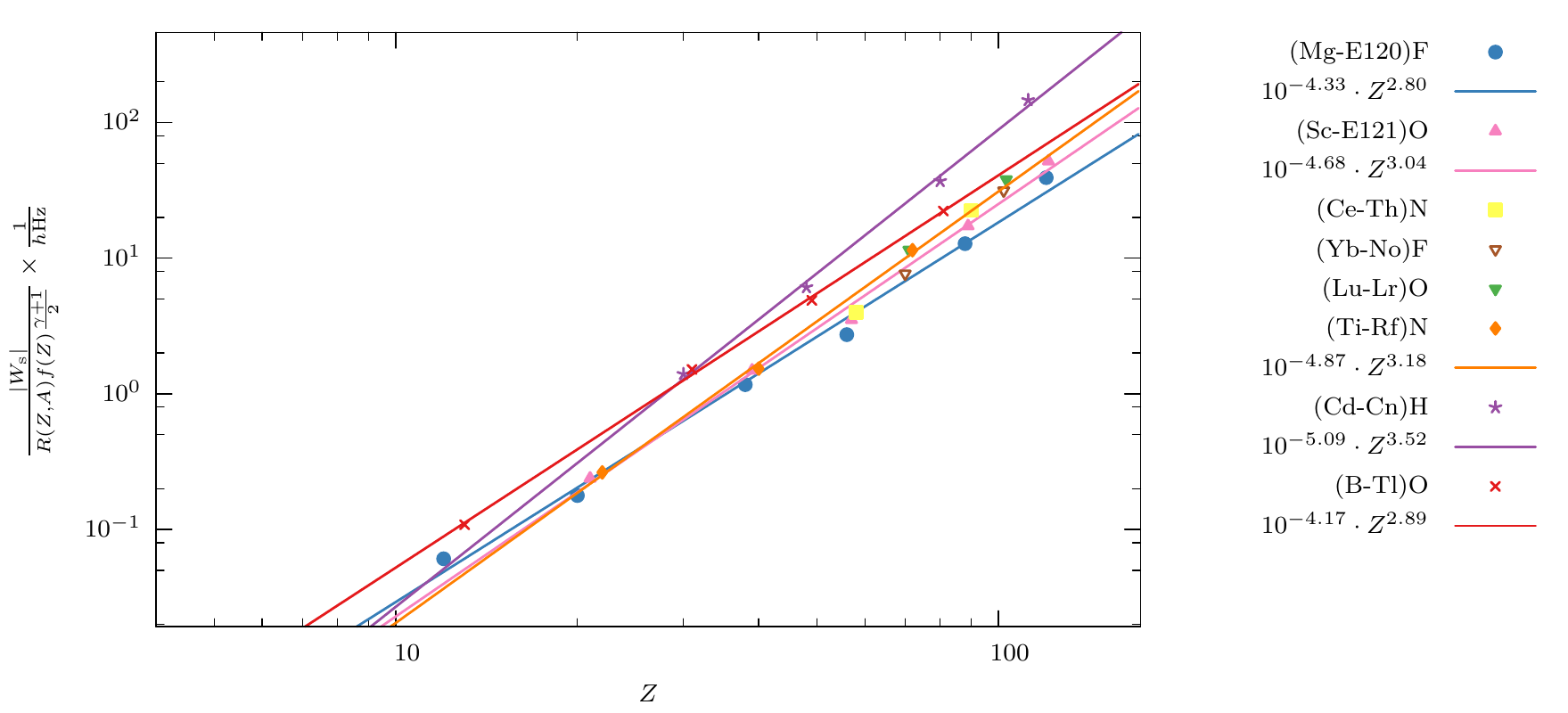}}
\resizebox{\textwidth}{!}{\includegraphics{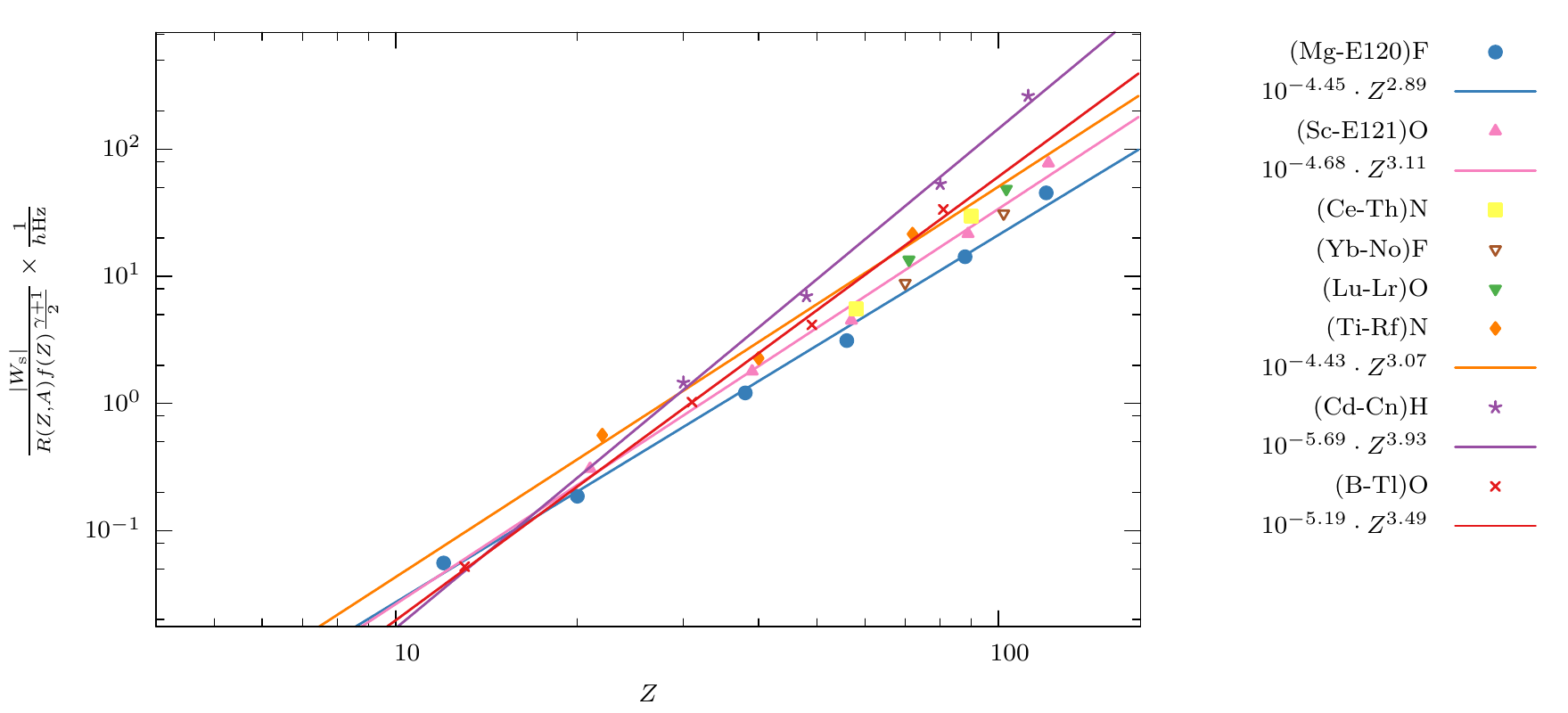}}
\caption{
Scaling of
$\log_{10}\braces{\frac{|W_\text{s}|}{R(Z,A)f(Z)\frac{\gamma+1}{2}}\times\frac{1}{h\mathrm{Hz}}}$ 
with $\log_{10}\braces{Z}$ for group 2 fluorides (Mg-E120)F, group 3
oxides (Sc-E121)O, group 4 nitrides (Ti-Hf)N, group 12 hydrides (Zn-Cn)H
and group 13 oxides (B-Tl)O at the level of GKS-ZORA/B3LYP (top) and
GHF-ZORA (bottom). The
functional expressions of the fits are assigned to the colors of the
groups. Plot of the $f$-block groups (Ce-Th)N, (Yb-No)F and (Lu-Lr)O
without fit. Boron was not included in the fit of group 13 oxides (see
text).}
\label{fig: wstrendsgks}
\end{figure*}
\begin{table*}\centering
\sisetup{table-number-alignment=center,table-figures-integer=3
,table-figures-decimal =2}
\begin{threeparttable}
\caption{$Z$-scaling $a$ and $Z$-independent factors $b$ of
$\frac{|W_\text{s}|}{R(Z,A)f(Z)\frac{\gamma+1}{2}}$ and $|W_\text{d}|\gamma^4$ (empirical
relativistic enhancement factor) for group 2
fluorides (Mg-Ra)F, group 3 oxides (Sc-Ac)O, group 4 nitrides
(Ti-Hf)N, group 12 hydrides (Zn-Cn)H and group 13 oxides (Al-Tl)O  at
the level of GHF-ZORA and GKS-ZORA/B3LYP.}
\label{tab: cpvtrends}
\begin{tabular}{l
S[separate-uncertainty,table-figures-uncertainty=1]
S[separate-uncertainty,table-figures-uncertainty=1]
c
S[separate-uncertainty,table-figures-uncertainty=1]
S[separate-uncertainty,table-figures-uncertainty=1]
c
S[separate-uncertainty,table-figures-uncertainty=1]
S[separate-uncertainty,table-figures-uncertainty=1]
c
S[separate-uncertainty,table-figures-uncertainty=1]
S[separate-uncertainty,table-figures-uncertainty=1]}
\toprule
\multirow{2}{*}{Group}&
\multicolumn{2}{c}{$a_\text{s}$}&&
\multicolumn{2}{c}{$b_\text{s}$}&&
\multicolumn{2}{c}{$a_\text{d,FS}$}&&
\multicolumn{2}{c}{$b_\text{d,FS}$}\\
\cline{2-3}\cline{5-6}\cline{8-9}\cline{11-12}
&\multicolumn{1}{c}{GHF}&\multicolumn{1}{c}{GKS}&
&\multicolumn{1}{c}{GHF}&\multicolumn{1}{c}{GKS}&
&\multicolumn{1}{c}{GHF}&\multicolumn{1}{c}{GKS}&
&\multicolumn{1}{c}{GHF}&\multicolumn{1}{c}{GKS}\\
\midrule                 
(Mg-E120)F
&2.88\pm0.10&2.79\pm0.11&&-4.44\pm0.16&-4.33\pm0.19&
&2.67\pm0.06&2.57\pm0.08&&-4.27\pm0.10&-4.14\pm0.13\\
(Sc-E121)O
&3.10\pm0.16&3.03\pm0.13&&-4.6\pm0.2&-4.6\pm0.2&
&2.81\pm0.07&2.75\pm0.10&&-4.36\pm0.12&-4.36\pm0.18\\
(Ti-Hf)N
&3.0\pm0.4&3.17\pm0.13&&-4.4\pm0.6&-4.8\pm0.2&
&3.0\pm0.4&3.16\pm0.12&&-4.5\pm0.6&-4.9\pm0.2\\
(Cd-Cn)H
&3.92\pm0.19&3.51\pm0.13&&-5.6\pm0.3&-5.0\pm0.2&
&3.72\pm0.11&3.31\pm0.05&&-5.5\pm0.2&-4.91\pm0.08\\
(Al-Tl)O
&3.48\pm0.12&2.89\pm0.05&&-5.1\pm0.2&-4.17\pm0.09&
&3.77\pm0.12&2.86\pm0.06&&-5.86\pm0.19&-4.27\pm0.08\\
\bottomrule
\end{tabular}
\end{threeparttable}
\end{table*}

\begin{figure*}[h]\centering
\resizebox{\textwidth}{!}{\includegraphics{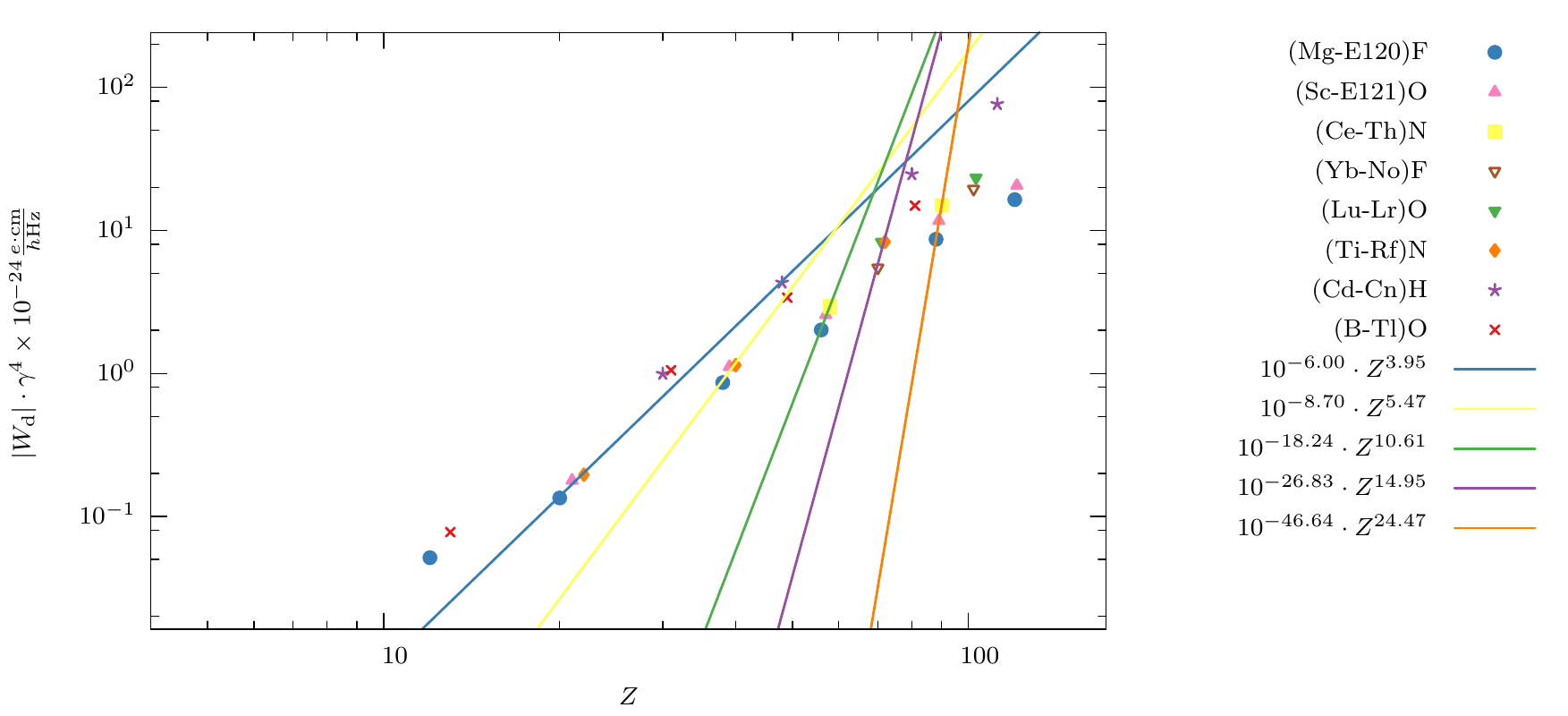}}
\resizebox{\textwidth}{!}{\includegraphics{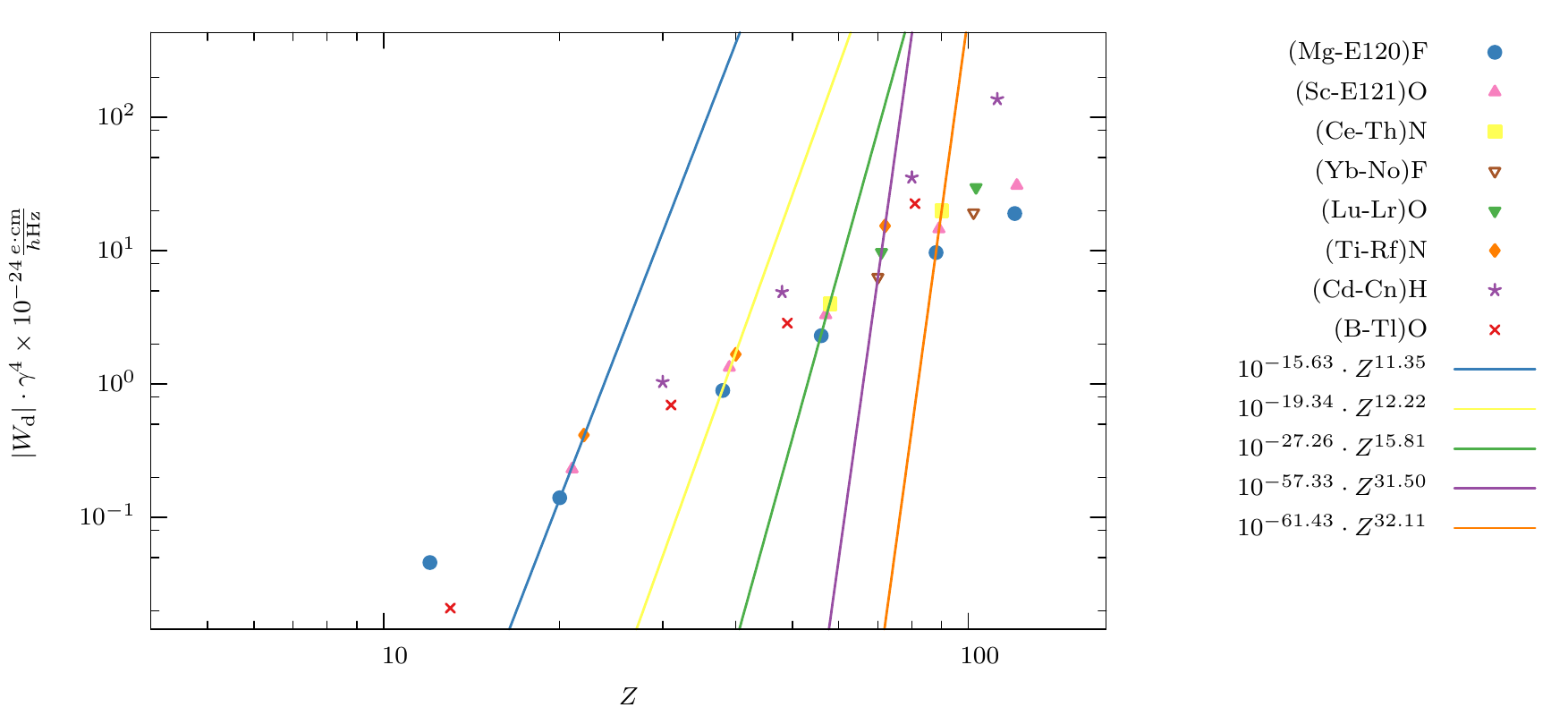}}
\caption{
Scaling of $\log_{10}\braces{|W_\text{d}|\gamma^4\times10^{-24}~\frac{e\cdot\mathrm{cm}}{h\mathrm{Hz}}}$  
with $\log_{10}\braces{Z}$ for row 4 (Ca-Ti; blue line), row 5
(Sr-Zr; yellow line), row 6 (Ba-Ce; green line, Yb-Hf; violet line),
and row 7 (Ra-Th; organge line, No-Lr; red line) at the level of
GKS-ZORA/B3LYP (top) and
GHF-ZORA (bottom). }
\label{fig: wdrowtrendsgks}
\end{figure*}
\begin{table*}\centering
\sisetup{table-number-alignment=center,table-figures-integer=3
,table-figures-decimal =2}
\begin{threeparttable}
\caption{$Z$-scaling $a$ and $Z$-indendent factors $b$ of
$\frac{|W_\text{s}|}{R(Z,A)f(Z)\frac{\gamma+1}{2}}$ and $|W_\text{d}|\gamma^4$ for isolobal diatomic molecules in row 4 (Ca-Ti), row 5 (Sr-Zr), row 6 (Ba-Ce; Yb-Hf),  and row 7 (Ra-Th; No-Lr)  at the level of  GHF/GKS-ZORA.}
\label{tab: cpvrowtrends}
{
\begin{tabular}{l
S[separate-uncertainty,table-figures-uncertainty=1]
S[separate-uncertainty,table-figures-uncertainty=1]
c
S[separate-uncertainty,table-figures-uncertainty=1]
S[separate-uncertainty,table-figures-uncertainty=1]
c
S[separate-uncertainty,table-figures-uncertainty=1]
S[separate-uncertainty,table-figures-uncertainty=1]
c
S[separate-uncertainty,table-figures-uncertainty=1]
S[separate-uncertainty,table-figures-uncertainty=1]}
\toprule
\multirow{2}{*}{Row}&
\multicolumn{2}{c}{$a_\text{s}$}&&
\multicolumn{2}{c}{$b_\text{s}$}&&
\multicolumn{2}{c}{$a_\text{d,FS}$}&&
\multicolumn{2}{c}{$b_\text{d,FS}$}\\
\cline{2-3}\cline{5-6}\cline{8-9}\cline{11-12}
&\multicolumn{1}{c}{GHF}&\multicolumn{1}{c}{GKS}
&                        
&\multicolumn{1}{c}{GHF}&\multicolumn{1}{c}{GKS}
&                        
&\multicolumn{1}{c}{GHF}&\multicolumn{1}{c}{GKS}
&                        
&\multicolumn{1}{c}{GHF}&\multicolumn{1}{c}{GKS}\\
\midrule
4 (Ca-Ti)
&11.5\pm0.8& 4.1\pm1.0&
&- 15.8\pm1.0&-6.1\pm1.4&

&11.3\pm0.7&3.9\pm1.1&
&- 15.6\pm0.9&- 6.0\pm 1.4\\
5 (Sr-Zr)
&12.2\pm1.6& 5\pm2&
&- 19\pm2&-8\pm3&

&12.2\pm1.8& 5\pm2&
&- 19\pm3&- 8\pm 3\\
6 (Ba-Ce)
&16\pm2&10\pm2&
&-28\pm4&-18\pm3&

&15\pm2&10.6\pm1.8&
&- 27\pm4&-18\pm 3\\
6 (Yb-Hf)
&31.6\pm1.0&14\pm8&
&-57.3\pm1.8&-25\pm15&

&31.5\pm0.6&14\pm8&
&-57.3\pm1.2&-26\pm15\\
7 (Ra-Th)
&33\pm2&25.1\pm0.9&
&-62\pm4&-47.7\pm1.8&

&32\pm2&24.4\pm1.0&
&- 61\pm4&-46\pm 2\\
\bottomrule
\end{tabular}
}
\end{threeparttable}
\end{table*}

\end{document}